\documentclass[notitlepage,aps,prl,reprint,twocolumn,longbibliography,superscriptaddress]{revtex4-1}
\usepackage{graphicx}
\usepackage{amsmath}
\usepackage{amssymb}
\usepackage{comment}
\usepackage[colorlinks, allcolors=blue]{hyperref}
\usepackage[all]{hypcap}
\usepackage[mathlines]{lineno}
\usepackage{braket}
\usepackage{wrapfig}
\usepackage{lipsum}
\usepackage{ulem}

\newcommand{\pref}[2]{\hyperref[#1]{\ref{#1}(#2)}}
\newcommand{\preff}[2]{\hyperref[#1]{\ref{#1 b}#2}}
\newcommand{\eqpref}[1]{\hyperref[#1]{(\ref{#1})}}

\newcommand{\squig}{{\raise.17ex\hbox{$\scriptstyle\sim$}}}

\begin{document}
\title{Non-Abelian Gauge Field Mechanics}
\author{Ivan Velkovsky}
\thanks{These authors contributed equally to this work.}
\affiliation{Department of Physics, University of Illinois at Urbana-Champaign, Urbana, IL 61801-3080, USA}
\author{Carlos Camacho}
\thanks{These authors contributed equally to this work.}
\affiliation{School of Physics and Astronomy, University of Birmingham, Edgbaston, Birmingham B15 2TT, United Kingdom}
\author{Tomoki Ozawa}
\email{tomoki.ozawa.d8@tohoku.ac.jp}
\affiliation{Advanced Institute for Materials Research (WPI-AIMR), Tohoku University, Sendai 980-8577, Japan}
\affiliation{RIKEN Center for Interdisciplinary Theoretical and Mathematical Sciences (iTHEMS), RIKEN, Wako, Saitama 351-0198, Japan}
\author{Hannah Price}
\email{H.Price.2@bham.ac.uk}
\affiliation{School of Physics and Astronomy, University of Birmingham, Edgbaston, Birmingham B15 2TT, United Kingdom}
\author{Bryce Gadway}
\email{bgadway@psu.edu}
\affiliation{Department of Physics, University of Illinois at Urbana-Champaign, Urbana, IL 61801-3080, USA}
\date{\today}
\affiliation{Department of Physics, The Pennsylvania State University, University Park, Pennsylvania 16802, USA}

\begin{abstract}
Non-Abelian gauge fields play a key role in describing the behavior of particles whose motion is coupled to internal degrees of freedom, such as their spin. Here, we experimentally realize a tuneable non-Abelian gauge field in an active mechanical lattice by using pairs of oscillators to encode a local pseudo-spin for each site, with inter-site spin-dependent couplings engineered via real-time measurement and feedback. We experimentally extract Wilson-loop observables in our set-up and hence demonstrate that we can create a genuinely non-Abelian gauge field. We then exploit the controllability of our mechanical lattice to engineer non-reciprocal hoppings to explore non-Hermitian non-Abelian gauge potentials. For a two-dimensional (2D) lattice, we demonstrate that the non-Hermiticity can manifest in direction-dependent Wilson loops for a single plaquette, while for a one-dimensional (1D) system, we show that a non-Abelian gauge potential can switch the localization of non-Hermitian skin modes between opposite ends of a chain. Our work establishes active mechanical lattices as a flexible and programmable platform for probing non-Abelian gauge fields and exploring their interplay with non-Hermitian dynamics.
\end{abstract}

\maketitle

Non-Abelian gauge fields underlie a wide range of phenomena across condensed matter physics~\cite{demler2004so,nayak2008non}, particle physics~\cite{yang1954conservation}, atomic physics~\cite{osterloh2005cold,Ohberg-NA,wu2016realization,huang2016experimental,sugawa2018second,GoldmanNA, goldman2014light,Liang2024} and photonics~\cite{yang2024non, cheng2023artificial,Zhang2022,cheng2025non}. One familiar example is spin-orbit coupling, where a particle’s spin and motion are linked together as effectively described by a non-Abelian gauge potential. A key feature of such a potential is that transporting a particle around a closed path produces a matrix-valued geometric transformation, or holonomy, acting on the particle's internal state. When the associated field is genuinely non-Abelian, some holonomies do not commute, meaning that the particle's internal state depends on the order of different paths being traversed~\cite{wu1975concept}. This contrasts with transport in an Abelian
field in which closed paths generate commuting (Aharonov–Bohm) phase factors~\cite{aharonov1959significance}.

The richness of non-Abelian gauge fields has motivated extensive efforts to realize such fields synthetically in controllable atomic~\cite{osterloh2005cold,Ohberg-NA,wu2016realization,huang2016experimental,sugawa2018second,GoldmanNA, goldman2014light,Liang2024}, acoustic~\cite{yang2024non}, optomechanical~\cite{Patil2022}, condensed matter~\cite{Wang2025nonmatter}, electromagnetic~\cite{Guo2021nonem} and photonic~\cite{yang2019synthesis, SoljacicNA,cheng2023artificial,Zhang2022,cheng2025non, yan2023non} platforms. In these set-ups, the role of the internal state can be played, e.g., by different hyperfine states~\cite{huang2016experimental,wu2016realization}, modes~\cite{yang2019synthesis}, or polarization states~\cite{cheng2023artificial,cheng2025non}, while the gauge field is realized by engineering couplings to imprint state-dependent phases and rotations during hopping processes. These approaches have been exploited to realize synthetic spin-orbit coupling~\cite{huang2016experimental, wu2016realization}, non-Abelian monopoles~\cite{sugawa2018second} and non-Abelian Aharonov–Bohm effects~\cite{yang2019synthesis} amongst other phenomena. 

Our work extends this physics into a classical mechanical setting by using measurement-based feedback to directly engineer a fully programmable non-dynamical gauge field for an array of classical oscillators~\cite{Anandwade-synthetic,Martello-Coexistence,Singhal-NH-AB,tian2023observation,velkovsky2024observation,rhyno}. To verify that this field can be genuinely non-Abelian, we use a protocol inspired by quantum process tomography to measure Wilson-loop observables, identifying when the internal state depends on the order in which different loops are traversed. Because our feedback-imposed couplings are not constrained by passive mechanical reciprocity, we exploit the same platform to explore the interplay of non-Abelian gauge potentials with non-Hermitian physics~\cite{Kunst-NH-BulkBound,Ueda-NH-review, zhang2022review}. We experimentally demonstrate that, in a 2D non-Hermitian lattice, the Wilson loop becomes sensitive to the direction in which a single plaquette is traversed, while in a 1D non-Hermitian lattice, a non-Abelian gauge potential can switch the localization of modes from one end of a chain to the other, enriching the physics of the non-Hermitian skin effect~\cite{Pang}. 

Our experiment consists of an array of feedback-coupled classical mechanical oscillators mapped onto an active mechanical lattice~\cite{Salerno_2014_epl, Salerno-floquet, Salerno_2017NJP, Brandenbourger2019,Ilan-prop,Top-Morph-active,Veenstra2024}, as detailed in Refs.~\cite{Anandwade-synthetic,Martello-Coexistence,Singhal-NH-AB,tian2023observation,velkovsky2024observation,rhyno}. In this mapping, every lattice site is represented by two oscillators, whose relative amplitude and phase encode a local SU(2) pseudo-spin degree of freedom. Physically, all the oscillators are uncoupled such that there is no intrinsic exchange of energy (phonons) between them. Instead, we use real-time measurements and feedback~\cite{Anandwade-synthetic} to encode the desired couplings, allowing energy (phonons) to “hop” between oscillators. Importantly, these couplings are not constrained to mimic those of physical springs, and can instead encode (non-Abelian) gauge fields and non-reciprocal (i.e. non-Hermitian) effects, as we explore in this paper. Although our experiments operate deep in the classical regime, they allow us to study the fundamental effects of non-Abelian gauge potentials on lattice physics. 

{\it 2D lattice with a genuine non-Abelian gauge field:} To demonstrate a tuneable non-Abelian gauge field, we consider a spin-1/2 particle hopping on a square lattice with a Hamiltonian~\cite{GoldmanNA}
\begin{equation}
H= J  \sum_{m,n} ( \hat{c}^\dagger_{m+1, n}  {X}  
\hat{c}_{m,n} + \hat{c}^\dagger_{m, n+1} {Y}  \hat{c}_{m,n} ) + \text{h.c.}, \label{eq:H1}
\end{equation}
where $ \hat{c}_{m,n}\! =\! (\hat{c}_{m,n}^\uparrow, \hat{c}_{m,n}^\downarrow)^T$ is the annihilation operator for the site indexed by $(m,n)$, $J$ is the hopping amplitude and ${X}$ and ${Y}$ are the spin-dependent tunneling operators, i.e. the link variables. Throughout we set $e\!=\!\hbar\!=\!1$. The tunneling operators can be expressed as ${X} \!=\! e^{i A_x}$ and ${Y} \!=\! e^{i A_y}$, with the gauge potential parameterized as
\begin{align}
	\mathbf{A} = (A_x, A_y)= (\alpha\sigma_{y}, \beta\sigma_{x}), 
	\label{eq:NAGauge}
\end{align}
where  $(\alpha, \beta)$ are real parameters and $\sigma_{x,y}$ are the Pauli matrices. 
Hence, the $X$ and $Y$ link variables generate rotations about the $y$ and $x$ pseudo-spin axes respectively as a particle hops. 

A non-Abelian gauge potential is often defined by $[A_x,A_y] \neq0$, i.e. that the gauge-potential components do not commute. However, it is important to note that this does not necessarily mean that the lattice gauge field (and hence the dynamics) is non-Abelian~\cite{goldman2014light}. For example, in Eq.~\ref{eq:NAGauge} if either $\alpha = d \pi $ or $\beta = d \pi $ where $d\in \mathbb{Z}$ then the link variable  $X$ or $Y$, respectively, will become equal to the identity, meaning that the system is Abelian despite $[A_x,A_y] \neq 0$. 

\begin{figure*}[t!]
	\includegraphics[width=1.75\columnwidth]{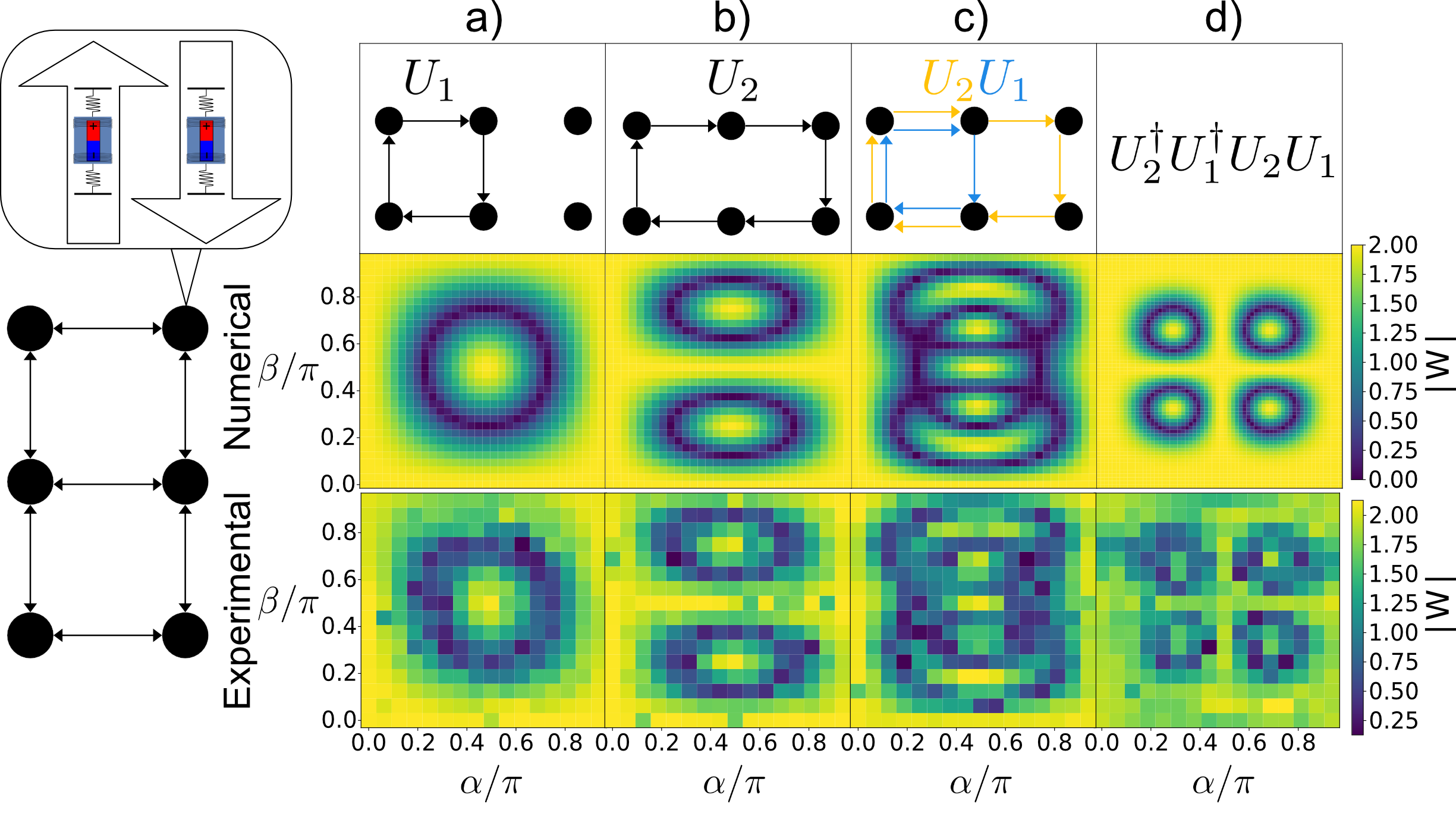}
	\centering
	\caption{\label{FIG:WL}
	{\it Left}: Schematic of the 6-site mechanical lattice. Two classical oscillators encode a local pseudo-spin-1/2 degree of freedom at each site. Each oscillator consists of a mass with a small dipole magnet suspended by springs between a pair of anti-Helmholtz coils. The current through these coils, and hence the force exerted on the masses, is determined dynamically through real-time feedback and measurement via an embedded accelerometer. Using these forces, we can map the hopping of phonons in our mechanical lattice to the dynamics of a particle moving under the Hamiltonian in Eq.~\ref{eq:H1}.  
    (a)-(d) Magnitude of the Wilson loop, $|W|$, for different choices of loop in our 6-site mechanical lattice. ({\it Top row}) Schematic of the measured loops corresponding to (a) $U_1$, (b) $U_2$, (c) $U_3=U_2 U_1$ and (d) $U_4= U_2^\dagger U_1^\dagger U_2 U_1$. Corresponding numerical calculations ({\it middle row}) and experimental measurements ({\it bottom row}) of $|W|$ as a function of parameters $(\alpha, \beta)$ [c.f. Eq.~\ref{eq:H1}]. The gauge field is genuinely non-Abelian when $|W|\neq 2$ for the ``commutator loop", $U_4$, in column (d). 
	}
\end{figure*}

To identify a {\it geninuely} non-Abelian gauge field, we must consider all the loop operators, $U_i$, for a system, where $i$ indexes all possible loops. For example, in the Abelian Harper-Hofstadter model, the simplest loop operator is for a single plaquette, $U_1\!=\!e^{i 2 \pi \phi}$ where $\phi$ is the number of (Abelian) magnetic flux quanta per plaquette, encoding the usual Aharonov-Bohm effect~\cite{aharonov1959significance}. By contrast, in our model, this operator becomes $U_1 = {Y}^\dagger {X}^\dagger {Y} {X}$ [c.f.~Fig.~\ref{FIG:WL}]; in the continuum limit, this can be related to the SU(2) Yang-Mills field strength and hence the non-Abelian Aharonov-Bohm effect~\cite{yang1954conservation,wu1975concept}. 

We must then consider what happens when a particle traverses successive loops starting from the same point, i.e. with $U = U_j U_i$ for some $i$ and $j$. One example is traversing a single-plaquette loop followed by a double-plaquette loop, i.e. $U=U_2 U_1$ with $U_2 = {Y}^\dagger {X}^\dagger {X}^\dagger  {Y} {X} {X}$ [c.f.~Fig.~\ref{FIG:WL}]. The net effect of any such process is that the particle acquires a geometric phase matrix as $U_j, U_i \in SU(N)$. If this is {\it not} the same phase matrix as occurs when traversing the loops in the opposite order (i.e. $U_j U_i \neq U_i U_j$), then the gauge field is genuinely non-Abelian~\cite{goldman2014light,SoljacicNA}. Note that if it is the same phase matrix, we cannot infer that the gauge field is Abelian; instead, $[U_i, U_j]=0$ must be confirmed for all $ i, j$, i.e. for every pair of loops in the system.

In our experiment, we first implement a non-Abelian gauge field using Eq.~\ref{eq:H1}. To implement this model, we employ a small lattice with two unit cells and a total of six sites, corresponding to twelve physical oscillators, as shown schematically in Fig.~\ref{FIG:WL}. To realize the gauge potential in Eq.~\ref{eq:NAGauge}, we use real-time measurement and feedback to impose the required coupling amplitudes and phases between the pairs of oscillators representing neighboring sites~\cite{Anandwade-synthetic}. The applied inter-oscillator forces simply follow Hamilton's equations of the desired model Hamiltonian, Eq.~\ref{eq:H1}.
The implemented gauge field is highly tuneable, with $\alpha$ and $\beta$ chosen freely in experiment.
We validate the implementation of Eq.~\ref{eq:H1} by measurement of its energy spectrum, as shown in the Supplemental Material~\cite{SuppMats}.

As our lattice is small, all possible loops can be decomposed in a basis of just four loop operators, $U_1$, $U_1^\dagger$, $U_2$, $U_2^\dagger$~\cite{SuppMats}. However, such loop operators are not generically invariant under local gauge transformations~\cite{goldman2014light}, reflecting that the Yang-Mills field strength is also gauge covariant. This motivates the introduction of the Wilson loop~\cite{wilson1974confinement}, 
\begin{equation}
 W_i = \text{Tr} (U_i)  
\end{equation}
which is a gauge-invariant quantity defined for each loop. Note that for this model [c.f. Eq.~\ref{eq:H1}], the Wilson loops do not depend on the initial position in the lattice as all loop operators are independent of position~\cite{GoldmanNA}.  

Experimentally, we can extract the Wilson loop through a sequence of four measurements in a procedure inspired by quantum process tomography~\cite{chuang1997prescription}. In this procedure, we initialize the system by injecting energy into a single
lattice site in one of four states
\begin{equation}
 \ket{0} \equiv \ket{\uparrow},
  \ket{1} \equiv \ket{\downarrow},
  \ket{2} \equiv \frac{(\ket{\uparrow} + \ket{\downarrow})}{\sqrt{2}},
  \ket{3} \equiv \frac{(\ket{\uparrow} - i\ket{\downarrow})}{\sqrt{2}}, \nonumber
 \label{eq:NAStates}
\end{equation}
by exciting the two oscillators of that
site with a suitable phase relationship. We then turn
on a single coupling link for exactly one tunneling time,
calibrated against our empirically-measured experimental tunneling rate. This coupling transfers the state to
the next site while performing the desired rotation in the
pseudo-spin subspace [c.f. Eq.~\ref{eq:H1}]. We repeat this procedure for each of the bonds within a given loop until the excitation returns to the original site. By comparing the final states obtained from the chosen four initial states, we are able to extract the magnitude of the corresponding Wilson loop, $|W_i|$, as detailed in the Supplemental Material~\cite{SuppMats}. 

This procedure is applied to measure the magnitude of the Wilson loop for varying $\alpha$ and $\beta$ [c.f. Eq.~\ref{eq:H1}] as shown in Fig.~\ref{FIG:WL} for loops (a) $U_1$, (b) $U_2$, (c) $U_3=U_2 U_1$ and (d) $U_4= U_2^\dagger U_1^\dagger U_2 U_1$. We find good agreement between numerical simulations ({\it middle row}) and experimental measurements ({\it bottom row}). Sources of
error in our experiment are incomplete transfer of energy
during a single rotation as well as accumulated differences
in energy loss in each of the two oscillators representing
the spin. From these measurements, we can infer information about the gauge field. For example, it can be shown that if $|W_i| = 2$ for an SU(2) gauge field, then the corresponding loop operator must be proportional to the identity and hence $U_i$ (and $U_i^\dagger$) must commute with all other loop operators~\cite{goldman2014light}. For our finite-size system, it follows that the gauge field must be Abelian whenever $|W_i| = 2$ in panels (a)-(c) of Fig.~\ref{FIG:WL}~\cite{SuppMats}. From this, we can conclude that the gauge field is, in fact, Abelian for a wide-range of parameter values, even though $[A_x,A_y] \neq 0$ if $\alpha, \beta \neq 0$. 

Inferring that the gauge field is {\it geninuely non-Abelian} requires more care as measuring $|W_i|\! \neq \! 2$ is usually a necessary but not sufficient condition~\cite{goldman2014light}. This can be seen by comparing panels Fig.~\ref{FIG:WL}~(a)-(c), where we often measure $|W_i| \!\neq\! 2$ for one loop but $|W_i|\! =\! 2$ for another. Instead, we consider a special loop $U_4= U_2^\dagger U_1^\dagger U_2 U_1$ in panel (d), which directly probes the non-commutativity of the loop operators. For this loop, it can be shown that if $|W_4|\! \neq \! 2$ then the gauge field must be non-Abelian~\cite{SuppMats}. Hence, we can infer parameters for which our model has a geninuely non-Abelian gauge field. 

{\it Non-Hermitian Wilson Loops:} Non-Hermiticity enriches the physics of Wilson loops by relaxing the constraint of reciprocity present in Hermitian lattice models.
In a Hermitian system, link variables associated with opposite directions must be related by Hermitian conjugation, which forces clockwise (CW) and counter-clockwise (CCW) loop operators to obey $U_{CW}\!=\!U_{CCW}^{\dagger}$. This in turn forces the corresponding Wilson loops to satisfy $W_{CW}\!=\!W_{CCW}^*$, for both Abelian and non-Abelian gauge fields. Once this Hermitian constraint is relaxed, the Wilson loop can become sensitive to the ordering of non-commuting link variables, allowing for a
gauge-invariant witness of non-Abelian path ordering, as we now demonstrate.

We consider a model for a spin-1/2 particle on a square lattice with the Hamiltonian~\cite{Pang}
\begin{eqnarray}
H=& J  \sum_{m,n} [  \hat{c}^\dagger_{m+1, n}  e^{i \theta_R \sigma_x}
\hat{c}_{m,n} + \hat{c}^\dagger_{m, n+1} e^{i \theta_U \sigma_z} \hat{c}_{m,n} \nonumber \\
&+   \hat{c}^\dagger_{m, n}  e^{i \theta_L \sigma_y} 
\hat{c}_{m+1,n}  + \hat{c}^\dagger_{m, n} e^{i \theta_D {\mathbf{1}}} \hat{c}_{m,n+1}], \label{eq:H2}
\end{eqnarray}
where $\theta_R, \theta_U, \theta_L, \theta_D$ are real parameters associated with moving right/up/left/down in the lattice respectively. Physically, the corresponding link variables implement spin rotations about the $x$-, $z$-, and $y$-axes and a spin-independent $U(1)$ phase, respectively. Although all the individual link variables are unitary, the Hamiltonian is non-Hermitian because the link operator in one direction is not constrained to be the Hermitian conjugate of that in the opposite direction. Although such a model would be difficult to realize with a real quantum system, this is straightforward for us to engineer using measurement and feedback in our classical-mechanical lattice~\cite{Anandwade-synthetic}. 

\begin{figure}[t!]	\includegraphics[width=0.95\columnwidth]{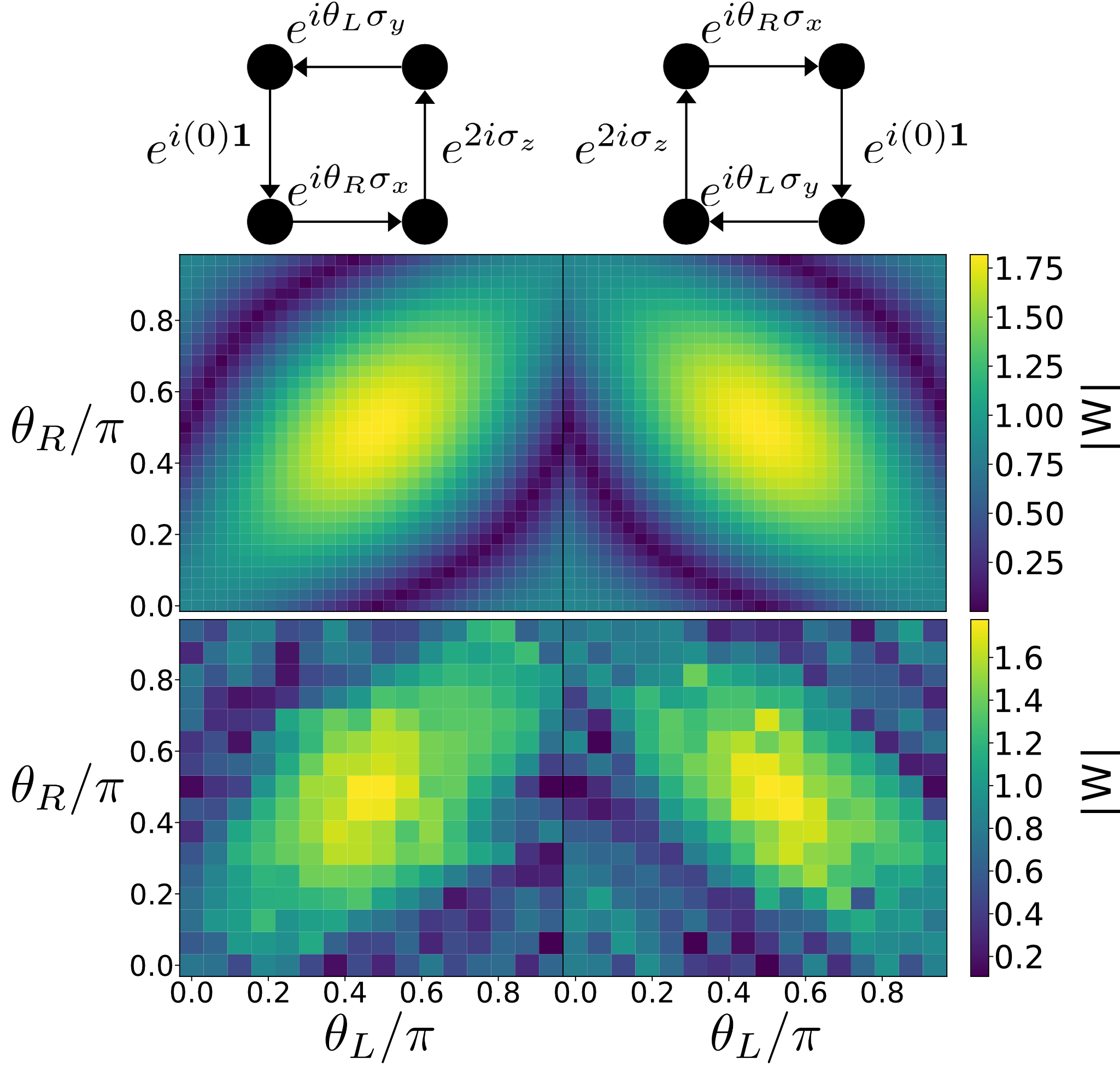}
	\centering
	\caption{\label{FIG:WL-NH}
		Magnitude of the Wilson loop, $|W|$, for (a) counter-clockwise and (b) clockwise single-plaquette loop operators, as indicated schematically ({\it top row}). Corresponding numerical calculations ({\it middle row}) and experimental measurements ({\it bottom row}) as a function of $\theta_R$ and $\theta_L$ with $\theta_D=0$, $\theta_U=2$ [c.f Eq.~\ref{eq:H2}]. Non-Hermiticity breaks the symmetry between the clockwise and counter-clockwise Wilson loops, making this measurement orientation-dependent. 
	}
\end{figure}

For this model, the single-plaquette loop operators can be chosen as $U_{CW}\!=\! e^{i \theta_L \sigma_y} e^{i \theta_D {\mathbf{1}}} e^{i \theta_R \sigma_x} e^{i \theta_U \sigma_z} $ and $U_{CCW}\!=\! e^{i \theta_D {\mathbf{1}}} e^{i \theta_L \sigma_y} e^{i \theta_U \sigma_z} e^{i \theta_R \sigma_x}$, which are evidently not related by Hermitian conjugation, $U_{CW}\!\neq\! U_{CCW}^\dagger$ [c.f. Fig.~\ref{FIG:WL-NH}]. Hence, the Wilson loops need not satisfy $W_{CW}\!=\! W_{CCW}^*$. Explicitly, they are given by
$W_{CW /CCW}\!=\!2e^{i\theta_D}
(\cos\theta_L\cos\theta_R\cos\theta_U
\pm \sin\theta_L\sin\theta_R\sin\theta_U)$~\cite{Pang};
thus the CW/CCW contrast is controlled by $\theta_L$, $\theta_R$, $\theta_U$, which are associated with the
non-commuting Pauli rotations.

To show this experimentally, we use the above Wilson-loop measurement procedure to extract $|W_{CW}|$ and $|W_{CCW}|$ for a single-plaquette of our mechanical lattice. These results are shown in Fig.~\ref{FIG:WL-NH} as a function of $\theta_L$ and $\theta_R$ (with $\theta_U=2$, $\theta_D=0$) for both numerics and experiment. For this choice
of parameters, the Wilson loops are always purely real, so the observed difference between the CW and CCW results directly reflects the path-dependence of the Wilson loop. Our measurements therefore provide evidence for non-Abelian link variables over a wide range of parameter values.

\begin{figure}[t!]
	\includegraphics[width=1\columnwidth]{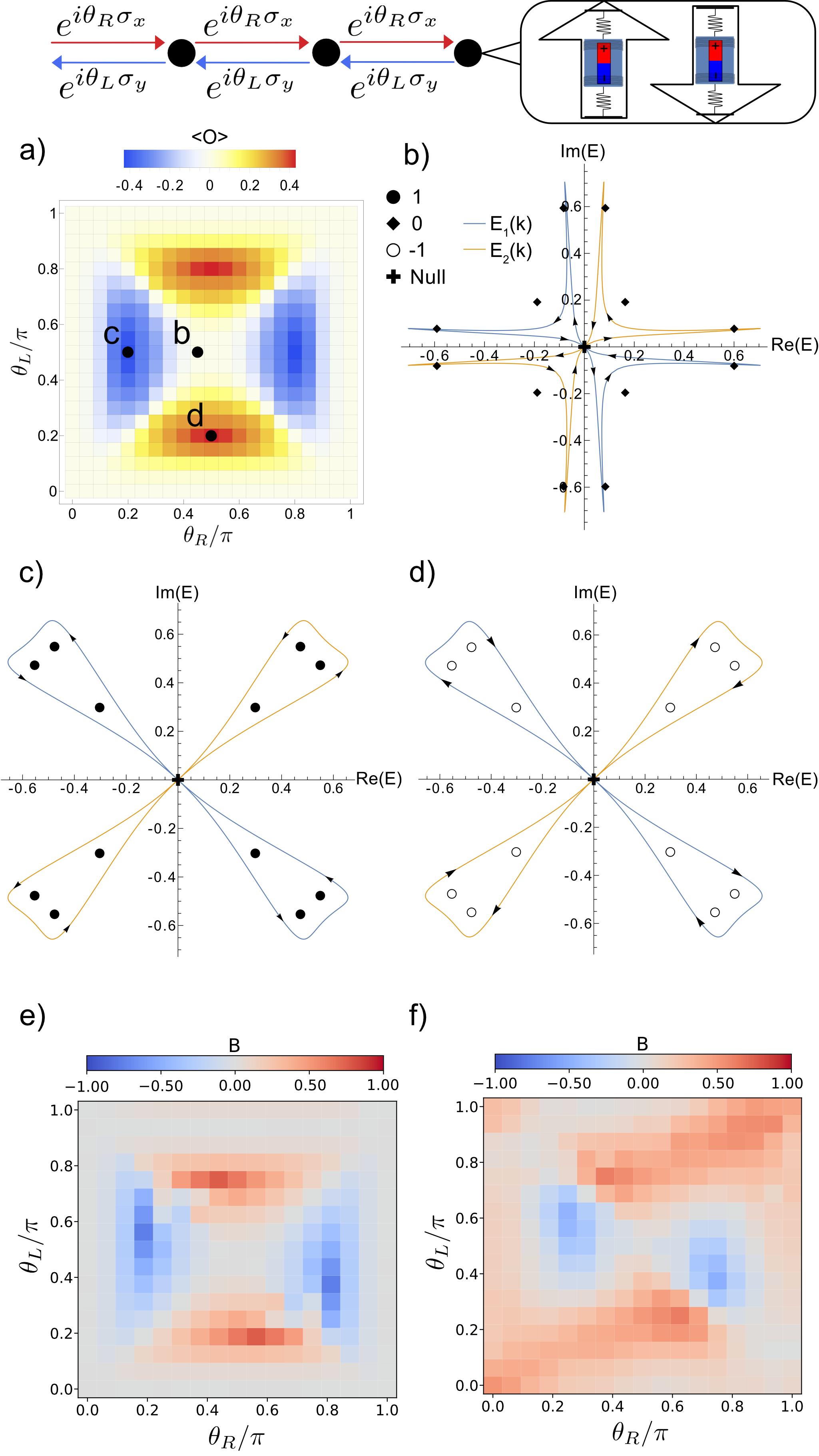}
	\centering
	\caption{\label{FIG:NH-Chain}
	({\it Top row}) Schematic of the 1D non-Abelian Hatano-Nelson model [Eq.~\ref{eq:HHN}]. Non-Hermiticity is introduced by hoppings to the left/right having different pseudo-spin rotations. (a) The left-right bias [Eq.~\ref{eq:bias}] for the OBC eigenstates of a 7-site chain, showing that $(\theta_R, \theta_L)$ can induce and switch the NHSE. (e) The numerical and (f) experimental left-right bias extracted from the oscillation amplitudes after excitation of the middle site and evolution for $T=60$~s, showing qualitative agreement. (b)-(d) The PBC spectrum [Eq.~\ref{eq:spectrum}] ({\it solid lines}) and the 7-site OBC eigenvalues ({\it symbols}), with parameters in (a). For each OBC eigenvalue, the symbol indicates if, with respect to that energy, the PBC spectrum: winds once counter-clockwise ({\it filled circle}), winds once clockwise ({\it empty circle}), does not wind ({\it diamond}) or has undefined winding ({\it cross}). Where there is no NHSE, e.g. in (a), the non-zero OBC eigenvalues do not have a net winding. For (e) and (f), where there is a strong NHSE, non-zero OBC eigenvalues have a counter-clockwise and clockwise winding, respectively. 
	}
\end{figure}

{\it 1D non-Abelian Hatano-Nelson Model:} In order to define a meaningful physical quantity due to a gauge potential, we require the existence of non-trivial closed loops in our lattice. For 1D Hermitian lattices, with open boundary conditions, there are no non-trivial closed loops from which to define our physical quantities. However, in 1D non-Hermitian systems, a non-Abelian gauge field can play a key role, as proposed in Ref.~\cite{Pang}. Following this work, we experimentally study a non-Abelian SU(2) variant of the 1D Hatano-Nelson model~\cite{hatano1996localization} given by
\begin{eqnarray}
H = \sum_m \left[J_L \hat{c}^\dagger_m e^{i \theta_L \sigma_y} \hat{c}_{m+1} + J_R \hat{c}^\dagger_{m+1} e^{i \theta_R \sigma_x} \hat{c}_{m} \right] \label{eq:HHN}
 \end{eqnarray}
where  $(J_{L}, J_R)$ are the non-reciprocal hopping amplitudes and $(\theta_{L}, \theta_{R})$ are the gauge-potential parameters. With periodic boundary conditions (PBC), this model has the complex-energy eigenspectrum~\cite{Pang},
\begin{eqnarray}
 E_\pm  =   A(k) \pm i \sqrt{J_L^2 \sin^2 \theta_L e^{2 i k} + J_R^2 \sin^2 \theta_R e^{-2 i k}}\label{eq:spectrum} 
\end{eqnarray}
where $A(k) = J_L \cos \theta_L e^{ik}+ J_R \cos \theta_R e^{-ik}$.

In the absence of the non-Abelian gauge potential ($\theta_R\!=\!\theta_L\!=\!0$), the Hatano-Nelson model~\cite{hatano1996localization} in Eq.~\ref{eq:HHN} is the prototypical model for the non-Hermitian skin effect (NHSE), in which a macroscopic number of eigenstates become localized at one end of an open chain~\cite{zhang2022review,lin2023topological}. Introducing the SU(2) gauge potential has been predicted to transform the NHSE, allowing for the switching of skin-mode localization from one end of the chain to the other by tuning the gauge-field angles $(\theta_R, \theta_L)$~\cite{Pang}. Hereafter, we set $J_L\!=J_R$ to focus on the non-Hermitian effects introduced solely by the gauge field. 

We probe the localization-switching of the NHSE experimentally by studying a chain of 7 spinful sites, corresponding to 14 physical oscillators. Numerically, we calculate the left-right bias along the chain, 
\begin{equation}
\mathcal{B} = \sum_{\alpha=1}^{2L} \sum_{m, \sigma} b_m |\psi^{(\alpha)}_{m, \sigma}|^2 \label{eq:bias}
\end{equation}
where $|\psi^{(\alpha)}_{m, \sigma}|^2$ is the normalized intensity for the open-boundary-condition (OBC) eigenstate indexed by $\alpha$, and $L$ is the number of lattice sites. For $L=7$, $b_m = (-1,-1,-1,0,1, 1, 1)$, such that $\mathcal{B}>0$ ($\mathcal{B}<0$) indicates that the eigenstates are predominantly localized on the right (left) side of the chain. This is shown in Fig.~\ref{FIG:NH-Chain}(a), where we see that varying $(\theta_{L}, \theta_{R})$ both induces the NHSE and switches it between opposite sides of the chain. Regions with opposite mode-localization are separated by the diagonals, $\theta_R \!=\! \theta_L$ and $\theta_R + \theta_L \!=\! \pi$, where the exceptional points occur and where modes extend across the chain~\cite{Pang}. The bias also vanishes around $\theta_R={0, \pi}$ and  $\theta_L={0, \pi}$, where the model [Eq.~\ref{eq:HHN}] reduces to a Hermitian 1D chain. 

The emergence of the NHSE can also be inferred from the point-gap topology of the PBC spectrum~\cite{zhang2022review,lin2023topological}. To see this, we plot the PBC spectrum [Eq.~\ref{eq:spectrum}] and OBC eigenvalues in the complex plane, as shown in Fig.~\ref{FIG:NH-Chain}  (b-d). As can be seen, the OBC eigenvalues deviate from the PBC spectrum, indicating the breakdown of conventional Bloch theory. Instead, the PBC eigenvalues form closed loops; in the absence of the NHSE, these loops do not wind around the OBC eigenvalues as in panel (b). However, when the NHSE is present, there is a clear winding around (non-zero) OBC eigenvalues [c.f. panels (c) and (d)], the sense of which switches with the localization. 

To observe this behavior in experiment, we initialize the system by injecting energy into a single oscillator in the middle of the chain at site ($m\!=\!4$, pseudospin $\ket{\downarrow}$), which generically has non-zero projection onto many of the system eigenmodes. We then examine the normalized population dynamics over a long period of $T = 60$~s. When the spectrum is complex [c.f.~Eq.~\ref{eq:spectrum}], such long-time dynamics are dominated by the relative growth of modes with positive imaginary energies, giving rise to the accumulation of population in boundary-localized skin modes. 
In analogy with the eigenstate bias [c.f.~Eq.~\ref{eq:bias}], we measure the difference between the sum of squared amplitudes (normalized oscillator energies) on the left and right halves of the chain at the end of the observation period $T$, defining the experimental ``bias''  $\mathcal{B}_\textrm{expt} = (\Sigma_{m < 4,\sigma} |\psi_{m,\sigma}|^2 - \Sigma_{m > 4,\sigma} |\psi_{m,\sigma}|^2) / (\Sigma_{m,\sigma} |\psi_{m,\sigma}|^2)$.
As Eq.~\ref{eq:HHN} is purely linear, we explore its spatial energy dynamics while stabilizing the \textit{total} energy via a weak measurement-based feedback that is decoupled from the underlying non-Hermitian physics~\cite{SuppMats}, so as to avoid the physical oscillators from experiencing large energy growth, significant nonlinearities, and potential damage.

Figure~\ref{FIG:NH-Chain} plots the long-time bias for (e) numerical simulations and (f) experimental data, respectively. These plots largely reproduce the main features of the theoretical eigenstate bias in panel (a), with the phases $\theta_{L/R}$ able to control and switch the appearance and imbalance of non-Hermitian skin modes in the system. The experiment systematically shows slightly more positive bias than the numerics, which we attribute to residual disparities in the damping corrections across the oscillator array~\cite{SuppMats,Anandwade-synthetic}.Not visible in the bias measurements, we additionally observe that $\theta_{L/R}$ control the frequency of local energy oscillations occurring during the dynamics, reflecting the concomitant change to the real parts of the eigenmode spectrum (c.f. panels (b-d) in Fig.~\ref{FIG:NH-Chain}).

{\it Conclusions:} We have experimentally realized a non-Abelian gauge field in an active mechanical lattice, using pairs of oscillators to encode a local pseudo-spin degree of freedom. The associated gauge field is fully tuneable, thanks to the programmability of inter-site couplings. We exploit this control to measure the magnitude of Wilson loops and hence, identify parameter values for which the gauge field is geninuely non-Abelian. By making the inter-site couplings non-reciprocal, we then study the interplay of non-Abelian gauge potentials with non-Hermitian physics. Firstly, for a 2D lattice, we experimentally show that the Wilson loop of a single plaquette can be different when traversed in a clockwise or counter-clockwise sense, unlike in a Hermitian system. Secondly, we demonstrate that the localization of modes in the NHSE can be switched across a 1D chain by tuning a non-Abelian gauge potential. 

Looking forward, this platform offers a route towards realizing programmable mechanical lattices which are governed by reconfigurable non-Abelian gauge fields. The flexibility of this set-up allows us to add nonlinear terms and/or go beyond static externally-imposed gauge fields; for example, inspired by density-dependent gauge fields in ultracold atoms~\cite{goldman2014light,greschner2015density,gorg2019realization} and dynamical gauge fields in analogous classical emulation platforms~\cite{FredJlgt}, the programmed link variables could be made amplitude dependent, so that the mechanical wave modifies the gauge field through which it propagates. This would open a route to explore nonlinear non-Abelian holonomies, interaction-dependent Wilson loops, or gauge-potential-controlled skin effects where the localization depends on excitation strength.

\section{Acknowledgements}

This work is supported by the Royal Society via grants
URF\textbackslash R\textbackslash221004 and RGF\textbackslash{}EA\textbackslash{}180121 and by the Engineering and Physical Sciences Research Council [grant numbers EP/W016141/1, EP/Y01510X/1 and UKRI2226].
This material (I.~V. and B.~G.) is based upon work supported by the National Science Foundation under grants No.~1945031 and No.~2438226. 
I.~V. and B.~G. also acknowledge support from the AFOSR MURI program under agreement number FA9550-22-1-0339.
TO is supported by JSPS KAKENHI Grant No. JP24K00548, JST PRESTO Grant No. JPMJPR2353, JST PRESTO Convergence Research Grant No. JPMJCR26XA

\bibliographystyle{apsrev4-1}
\bibliography{NA-NH}

@article{GoldmanNA,
  title = {Ultracold atomic gases in non-Abelian gauge potentials: The case of constant Wilson loop},
  author = {Goldman, N. and Kubasiak, A. and Gaspard, P. and Lewenstein, M.},
  journal = {Phys. Rev. A},
  volume = {79},
  issue = {2},
  pages = {023624},
  numpages = {11},
  year = {2009},
  month = {Feb},
  publisher = {American Physical Society},
  doi = {10.1103/PhysRevA.79.023624},
  url = {https://link.aps.org/doi/10.1103/PhysRevA.79.023624}
}

@Article{SoljacicNA,
author={Yang, Yi
and Zhen, Bo
and Joannopoulos, John D.
and Solja{\v{c}}i{\'{c}}, Marin},
title={Non-Abelian generalizations of the Hofstadter model: spin--orbit-coupled butterfly pairs},
journal={Light: Science {\&} Applications},
year={2020},
month={Oct},
day={19},
volume={9},
number={1},
pages={177},
issn={2047-7538},
doi={10.1038/s41377-020-00384-7},
url={https://doi.org/10.1038/s41377-020-00384-7}
}

@article{QuantumProcessTomography,
author = {Isaac L. Chuang and M. A. Nielsen},
title = {Prescription for experimental determination of the dynamics of a quantum black box},
journal = {Journal of Modern Optics},
volume = {44},
number = {11-12},
pages = {2455--2467},
year = {1997},
publisher = {Taylor \& Francis},
doi = {10.1080/09500349708231894}
}

@article{Pang,
  title = {Synthetic Non-Abelian Gauge Fields for Non-Hermitian Systems},
  author = {Pang, Zehai and Wong, Bengy Tsz Tsun and Hu, Jinbing and Yang, Yi},
  journal = {Phys. Rev. Lett.},
  volume = {132},
  issue = {4},
  pages = {043804},
  numpages = {7},
  year = {2024},
  month = {Jan},
  publisher = {American Physical Society},
  doi = {10.1103/PhysRevLett.132.043804},
  url = {https://link.aps.org/doi/10.1103/PhysRevLett.132.043804}
}

@Article{Brandenbourger2019,
author={Brandenbourger, Martin
and Locsin, Xander
and Lerner, Edan
and Coulais, Corentin},
title={Non-reciprocal robotic metamaterials},
journal={Nature Communications},
year={2019},
month={Oct},
day={10},
volume={10},
number={1},
pages={4608},
issn={2041-1723},
doi={10.1038/s41467-019-12599-3},
url={https://doi.org/10.1038/s41467-019-12599-3}
}

@article{Martello-Coexistence,
  title = {Coexistence of stable and unstable population dynamics in a nonlinear non-Hermitian mechanical dimer},
  author = {Martello, Enrico and Singhal, Yaashnaa and Gadway, Bryce and Ozawa, Tomoki and Price, Hannah M.},
  journal = {Phys. Rev. E},
  volume = {107},
  issue = {6},
  pages = {064211},
  numpages = {14},
  year = {2023},
  month = {Jun},
  publisher = {American Physical Society},
  doi = {10.1103/PhysRevE.107.064211}
}

@article{Singhal-NH-AB,
  title = {Measuring the adiabatic non-Hermitian Berry phase in feedback-coupled oscillators},
  author = {Singhal, Yaashnaa and Martello, Enrico and Agrawal, Shraddha and Ozawa, Tomoki and Price, Hannah and Gadway, Bryce},
  journal = {Phys. Rev. Res.},
  volume = {5},
  issue = {3},
  pages = {L032026},
  numpages = {7},
  year = {2023},
  month = {Aug},
  publisher = {American Physical Society},
  doi = {10.1103/PhysRevResearch.5.L032026}
}

@article{Kunst-NH-BulkBound,
  title = {Biorthogonal Bulk-Boundary Correspondence in Non-Hermitian Systems},
  author = {Kunst, Flore K. and Edvardsson, Elisabet and Budich, Jan Carl and Bergholtz, Emil J.},
  journal = {Phys. Rev. Lett.},
  volume = {121},
  issue = {2},
  pages = {026808},
  numpages = {6},
  year = {2018},
  month = {Jul},
  publisher = {American Physical Society},
  doi = {10.1103/PhysRevLett.121.026808},
  url = {https://link.aps.org/doi/10.1103/PhysRevLett.121.026808}
}

@article{Ueda-NH-review,
author = {Yuto Ashida and Zongping Gong and Masahito Ueda},
title = {Non-Hermitian physics},
journal = {Advances in Physics},
volume = {69},
number = {3},
pages = {249-435},
year  = {2020},
publisher = {Taylor & Francis},
doi = {10.1080/00018732.2021.1876991},
URL = {https://doi.org/10.1080/00018732.2021.1876991}
}

@article{Salerno_2014_epl,
	doi = {10.1209/0295-5075/106/24002},
	url = {https://doi.org/10.1209/0295-5075/106/24002},
	year = 2014,
	month = {apr},
	publisher = {{IOP} Publishing},
	volume = {106},
	number = {2},
	pages = {24002},
	author = {Grazia Salerno and Iacopo Carusotto},
	title = {Dynamical decoupling and dynamical isolation in temporally modulated coupled pendulums},
	journal = {{EPL} (Europhysics Letters)}
}

@article{Salerno_2017NJP,
	doi = {10.1088/1367-2630/aa6c03},
	url = {https://doi.org/10.1088/1367-2630/aa6c03},
	year = 2017,
	month = {may},
	publisher = {{IOP} Publishing},
	volume = {19},
	number = {5},
	pages = {055001},
	author = {Grazia Salerno and Alice Berardo and Tomoki Ozawa and Hannah M Price and Ludovic Taxis and Nicola M Pugno and Iacopo Carusotto},
	title = {Spin{\textendash}orbit coupling in a hexagonal ring of pendula},
	journal = {New Journal of Physics}
}

@misc{SuppMats,
  note = {See Supplementary Material for more details on the theoretical mapping, experimental system, additional measurements, and the theoretical analysis.}
}

@article{Ilan-prop,
  title = {Non-Newtonian Topological Mechanical Metamaterials Using Feedback Control},
  author = {Sirota, Lea and Ilan, Roni and Shokef, Yair and Lahini, Yoav},
  journal = {Phys. Rev. Lett.},
  volume = {125},
  issue = {25},
  pages = {256802},
  numpages = {6},
  year = {2020},
  month = {Dec},
  publisher = {American Physical Society},
  doi = {10.1103/PhysRevLett.125.256802},
  url = {https://link.aps.org/doi/10.1103/PhysRevLett.125.256802}
}

@Article{Top-Morph-active,
author={Wang, Wei
and Wang, Xulong
and Ma, Guancong},
title={Non-Hermitian morphing of topological modes},
journal={Nature},
year={2022},
month={Aug},
day={01},
volume={608},
number={7921},
pages={50-55},
issn={1476-4687},
doi={10.1038/s41586-022-04929-1}
}

@Article{Veenstra2024,
author={Veenstra, Jonas
and Gamayun, Oleksandr
and Guo, Xiaofei
and Sarvi, Anahita
and Meinersen, Chris Ventura
and Coulais, Corentin},
title={Non-reciprocal topological solitons in active metamaterials},
journal={Nature},
year={2024},
month={Mar},
day={01},
volume={627},
number={8004},
pages={528-533},
issn={1476-4687},
doi={10.1038/s41586-024-07097-6}
}

@article{Salerno-floquet,
  title = {Floquet topological system based on frequency-modulated classical coupled harmonic oscillators},
  author = {Salerno, Grazia and Ozawa, Tomoki and Price, Hannah M. and Carusotto, Iacopo},
  journal = {Phys. Rev. B},
  volume = {93},
  issue = {8},
  pages = {085105},
  numpages = {14},
  year = {2016},
  month = {Feb},
  publisher = {American Physical Society},
  doi = {10.1103/PhysRevB.93.085105},
  url = {https://link.aps.org/doi/10.1103/PhysRevB.93.085105}
}

@article{FredJlgt,
  title = {Engineering a U(1) lattice gauge theory in classical electric circuits},
  author = {Riechert, Hannes and Halimeh, Jad C. and Kasper, Valentin and Bretheau, Landry and Zohar, Erez and Hauke, Philipp and Jendrzejewski, Fred},
  journal = {Phys. Rev. B},
  volume = {105},
  issue = {20},
  pages = {205141},
  numpages = {12},
  year = {2022},
  month = {May},
  publisher = {American Physical Society},
  doi = {10.1103/PhysRevB.105.205141},
  url = {https://link.aps.org/doi/10.1103/PhysRevB.105.205141}
}

@article{tian2023observation,
  title = {Manipulation of Weyl Points in Reciprocal and Nonreciprocal Mechanical Lattices},
  author = {Tian, Mingsheng and Velkovsky, Ivan and Chen, Tao and Sun, Fengxiao and He, Qiongyi and Gadway, Bryce},
  journal = {Phys. Rev. Lett.},
  volume = {132},
  issue = {12},
  pages = {126602},
  numpages = {6},
  year = {2024},
  month = {Mar},
  publisher = {American Physical Society},
  doi = {10.1103/PhysRevLett.132.126602}
}

@article{Anandwade-synthetic,
  title = {Synthetic mechanical lattices with synthetic interactions},
  author = {Anandwade, Ritika and Singhal, Yaashnaa and Paladugu, Sai Naga Manoj and Martello, Enrico and Castle, Michael and Agrawal, Shraddha and Carlson, Ellen and Battle-McDonald, Cait and Ozawa, Tomoki and Price, Hannah M. and Gadway, Bryce},
  journal = {Phys. Rev. A},
  volume = {108},
  issue = {1},
  pages = {012221},
  numpages = {14},
  year = {2023},
  month = {Jul},
  publisher = {American Physical Society},
  doi = {10.1103/PhysRevA.108.012221}
}

@Article{Liang2024,
author={Liang, Qian
and Dong, Zhaoli
and Pan, Jian-Song
and Wang, Hongru
and Li, Hang
and Yang, Zhaoju
and Yi, Wei
and Yan, Bo},
title={Chiral dynamics of ultracold atoms under a tunable SU(2) synthetic gauge field},
journal={Nature Physics},
year={2024},
month={Nov},
day={01},
volume={20},
number={11},
pages={1738-1743},
issn={1745-2481},
doi={10.1038/s41567-024-02644-4}
}

@Article{Patil2022,
author={Patil, Yogesh S. S.
and H{\"o}ller, Judith
and Henry, Parker A.
and Guria, Chitres
and Zhang, Yiming
and Jiang, Luyao
and Kralj, Nenad
and Read, Nicholas
and Harris, Jack G. E.},
title={Measuring the knot of non-Hermitian degeneracies and non-commuting braids},
journal={Nature},
year={2022},
month={Jul},
day={01},
volume={607},
number={7918},
pages={271-275},
issn={1476-4687},
doi={10.1038/s41586-022-04796-w}
}

@article{Ohberg-NA,
  title = {Non-Abelian Gauge Potentials for Ultracold Atoms with Degenerate Dark States},
  author = {Ruseckas, J. and Juzeli\ifmmode \bar{u}\else \={u}\fi{}nas, G. and \"Ohberg, P. and Fleischhauer, M.},
  journal = {Phys. Rev. Lett.},
  volume = {95},
  issue = {1},
  pages = {010404},
  numpages = {4},
  year = {2005},
  month = {Jun},
  publisher = {American Physical Society},
  doi = {10.1103/PhysRevLett.95.010404}
}

@Article{Zhang2022,
author={Zhang, Xu-Lin
and Yu, Feng
and Chen, Ze-Guo
and Tian, Zhen-Nan
and Chen, Qi-Dai
and Sun, Hong-Bo
and Ma, Guancong},
title={Non-Abelian braiding on photonic chips},
journal={Nature Photonics},
year={2022},
month={May},
day={01},
volume={16},
number={5},
pages={390-395},
issn={1749-4893},
doi={10.1038/s41566-022-00976-2}
}

@article{rhyno,
  title = {Mechanical cosmology: Simulating scalar fluctuations in expanding universes using synthetic mechanical lattices},
  author = {Rhyno, Brendan and Velkovsky, Ivan and Adshead, Peter and Gadway, Bryce and Vishveshwara, Smitha},
  journal = {Phys. Rev. Res.},
  volume = {7},
  issue = {2},
  pages = {L022004},
  numpages = {7},
  year = {2025},
  month = {Apr},
  publisher = {American Physical Society},
  doi = {10.1103/PhysRevResearch.7.L022004},
  url = {https://link.aps.org/doi/10.1103/PhysRevResearch.7.L022004}
}

@article{velkovsky2024observation,
  title={Observation of chiral solitary waves in a nonlinear Aharonov-Bohm ring},
  author={Velkovsky, Ivan and Abraham, Anya and Martello, Enrico and Yu, Jiarui and Singhal, Yaashnaa and Gonzalez, Antonio and Lewis, DaVonte and Price, Hannah and Ozawa, Tomoki and Gadway, Bryce},
  journal={arXiv preprint arXiv:2406.01732},
  year={2024}
}

@article{cheng2025non,
author={Cheng, Dali
and Wang, Kai
and Roques-Carmes, Charles
and Lustig, Eran
and Long, Olivia Y.
and Wang, Heming
and Fan, Shanhui},
title={Non-Abelian lattice gauge fields in photonic synthetic frequency dimensions},
journal={Nature},
year={2025},
month={Jan},
day={01},
volume={637},
number={8044},
pages={52-56},
issn={1476-4687},
doi={10.1038/s41586-024-08259-2}
}

@article{yang1954conservation,
title = {Conservation of Isotopic Spin and Isotopic Gauge Invariance},
  author = {Yang, C. N. and Mills, R. L.},
  journal = {Phys. Rev.},
  volume = {96},
  issue = {1},
  pages = {191--195},
  numpages = {0},
  year = {1954},
  month = {Oct},
  publisher = {American Physical Society},
  doi = {10.1103/PhysRev.96.191}
}

@article{demler2004so,
  title = {$\mathit{SO}(5)$ theory of antiferromagnetism and superconductivity},
  author = {Demler, Eugene and Hanke, Werner and Zhang, Shou-Cheng},
  journal = {Rev. Mod. Phys.},
  volume = {76},
  issue = {3},
  pages = {909--974},
  numpages = {0},
  year = {2004},
  month = {Nov},
  publisher = {American Physical Society},
  doi = {10.1103/RevModPhys.76.909}
}

@article{nayak2008non,
  title = {Non-Abelian anyons and topological quantum computation},
  author = {Nayak, Chetan and Simon, Steven H. and Stern, Ady and Freedman, Michael and Das Sarma, Sankar},
  journal = {Rev. Mod. Phys.},
  volume = {80},
  issue = {3},
  pages = {1083--1159},
  numpages = {0},
  year = {2008},
  month = {Sep},
  publisher = {American Physical Society},
  doi = {10.1103/RevModPhys.80.1083}
}

@article{cheng2023artificial,
 title = {Artificial Non-Abelian Lattice Gauge Fields for Photons in the Synthetic Frequency Dimension},
  author = {Cheng, Dali and Wang, Kai and Fan, Shanhui},
  journal = {Phys. Rev. Lett.},
  volume = {130},
  issue = {8},
  pages = {083601},
  numpages = {8},
  year = {2023},
  month = {Feb},
  publisher = {American Physical Society},
  doi = {10.1103/PhysRevLett.130.083601}
}

@article{yang2024non,
author = {Yi Yang  and Biao Yang  and Guancong Ma  and Jensen Li  and Shuang Zhang  and C. T. Chan },
title = {Non-Abelian physics in light and sound},
journal = {Science},
volume = {383},
number = {6685},
pages = {eadf9621},
year = {2024},
doi = {10.1126/science.adf9621}
}

@article{Guo2021nonem,
language = {eng},
number = {7862},
pages = {195-200},
publisher = {Nature Publishing Group UK},
title = {Experimental observation of non-Abelian topological charges and edge states},
volume = {594},
year = {2021},
author = {Guo, Qinghua and Jiang, Tianshu and Zhang, Ruo-Yang and Zhang, Lei and Zhang, Zhao-Qing and Yang, Biao and Zhang, Shuang and Chan, C. T.},
address = {London},
issn = {0028-0836},
journal = {Nature},
doi = {10.1038/s41586-021-03521-3},
copyright = {The Author(s), under exclusive licence to Springer Nature Limited 2021},
}

@article{Wang2025nonmatter,
language = {eng},
number = {7},
pages = {873-880},
publisher = {Nature Publishing Group UK},
title = {Non-Hermitian non-Abelian topological transition in the S = 1 electron spin system of a nitrogen vacancy centre in diamond},
volume = {20},
year = {2025},
author = {Wang, Yunhan and Wu, Yang and Ye, Xiangyu and Duan, Chang-Kui and Wang, Ya and Hu, Haiping and Rong, Xing and Du, Jiangfeng},
address = {London},
issn = {1748-3387},
journal = {Nature nanotechnology},
doi = {10.1038/s41565-025-01913-4},
copyright = {The Author(s), under exclusive licence to Springer Nature Limited 2025 Springer Nature or its licensor (e.g. a society or other partner) holds exclusive rights to this article under a publishing agreement with the author(s) or other rightsholder(s); author self-archiving of the accepted manuscript version of this article is solely governed by the terms of such publishing agreement and applicable law.},
}

@article{yan2023non,
author = {Qiuchen Yan and Zhihao Wang and Dongyi Wang and Rui Ma and Cuicui Lu and Guancong Ma and Xiaoyong Hu and Qihuang Gong},
journal = {Adv. Opt. Photon.},
number = {4},
pages = {907--976},
publisher = {Optica Publishing Group},
title = {Non-Abelian gauge field in optics},
volume = {15},
month = {Dec},
year = {2023},
doi = {10.1364/AOP.494544}
}

@article{yang2019synthesis,
author = {Yi Yang  and Chao Peng  and Di Zhu  and Hrvoje Buljan  and John D. Joannopoulos  and Bo Zhen  and Marin Soljačić },
title = {Synthesis and observation of non-Abelian gauge fields in real space},
journal = {Science},
volume = {365},
number = {6457},
pages = {1021-1025},
year = {2019},
doi = {10.1126/science.aay3183}
}

@article{aharonov1959significance,
title = {Significance of Electromagnetic Potentials in the Quantum Theory},
  author = {Aharonov, Y. and Bohm, D.},
  journal = {Phys. Rev.},
  volume = {115},
  issue = {3},
  pages = {485--491},
  numpages = {0},
  year = {1959},
  month = {Aug},
  publisher = {American Physical Society},
  doi = {10.1103/PhysRev.115.485}
}

@article{huang2016experimental,
author={Huang, Lianghui
and Meng, Zengming
and Wang, Pengjun
and Peng, Peng
and Zhang, Shao-Liang
and Chen, Liangchao
and Li, Donghao
and Zhou, Qi
and Zhang, Jing},
title={Experimental realization of two-dimensional synthetic spin--orbit coupling in ultracold Fermi gases},
journal={Nature Physics},
year={2016},
month={Jun},
day={01},
volume={12},
number={6},
pages={540-544},
issn={1745-2481},
doi={10.1038/nphys3672}
}

@article{wu2016realization,
author = {Zhan Wu  and Long Zhang  and Wei Sun  and Xiao-Tian Xu  and Bao-Zong Wang  and Si-Cong Ji  and Youjin Deng  and Shuai Chen  and Xiong-Jun Liu  and Jian-Wei Pan },
title = {Realization of two-dimensional spin-orbit coupling for Bose-Einstein condensates},
journal = {Science},
volume = {354},
number = {6308},
pages = {83-88},
year = {2016},
doi = {10.1126/science.aaf6689}
}

@article{sugawa2018second,
  author = {Seiji Sugawa  and Francisco Salces-Carcoba  and Abigail R. Perry  and Yuchen Yue  and I. B. Spielman },
title = {Second Chern number of a quantum-simulated non-Abelian Yang monopole},
journal = {Science},
volume = {360},
number = {6396},
pages = {1429-1434},
year = {2018},
doi = {10.1126/science.aam9031}
}

@article{wu1975concept,
  title = {Concept of nonintegrable phase factors and global formulation of gauge fields},
  author = {Wu, Tai Tsun and Yang, Chen Ning},
  journal = {Phys. Rev. D},
  volume = {12},
  issue = {12},
  pages = {3845--3857},
  numpages = {0},
  year = {1975},
  month = {Dec},
  publisher = {American Physical Society},
  doi = {10.1103/PhysRevD.12.3845}
}

@article{goldman2014light,
doi = {10.1088/0034-4885/77/12/126401},
year = {2014},
month = {nov},
publisher = {IOP Publishing},
volume = {77},
number = {12},
pages = {126401},
author = {Goldman, N and Juzeliūnas, G and Öhberg, P and Spielman, I B},
title = {Light-induced gauge fields for ultracold atoms},
journal = {Reports on Progress in Physics}
}

@article{wilson1974confinement,
  title = {Confinement of quarks},
  author = {Wilson, Kenneth G.},
  journal = {Phys. Rev. D},
  volume = {10},
  issue = {8},
  pages = {2445--2459},
  numpages = {0},
  year = {1974},
  month = {Oct},
  publisher = {American Physical Society},
  doi = {10.1103/PhysRevD.10.2445}
}

@article{chuang1997prescription,
author = {Isaac L. Chuang and M. A. Nielsen},
title = {Prescription for experimental determination of the dynamics of a quantum black box},
journal = {Journal of Modern Optics},
volume = {44},
number = {11-12},
pages = {2455--2467},
year = {1997},
publisher = {Taylor \& Francis},
doi = {10.1080/09500349708231894}
}

@article{hatano1996localization,
title = {Localization Transitions in Non-Hermitian Quantum Mechanics},
  author = {Hatano, Naomichi and Nelson, David R.},
  journal = {Phys. Rev. Lett.},
  volume = {77},
  issue = {3},
  pages = {570--573},
  numpages = {0},
  year = {1996},
  month = {Jul},
  publisher = {American Physical Society},
  doi = {10.1103/PhysRevLett.77.570}
}

@article{zhang2022review,
author = {Xiujuan Zhang and Tian Zhang and Ming-Hui Lu and Yan-Feng Chen},
title = {A review on non-Hermitian skin effect},
journal = {Advances in Physics: X},
volume = {7},
number = {1},
pages = {2109431},
year = {2022},
publisher = {Taylor \& Francis},
doi = {10.1080/23746149.2022.2109431}
}

@article{lin2023topological,
  author={Lin, Rijia
and Tai, Tommy
and Li, Linhu
and Lee, Ching Hua},
title={Topological non-Hermitian skin effect},
journal={Frontiers of Physics},
year={2023},
month={Jul},
day={03},
volume={18},
number={5},
pages={53605},
issn={2095-0470},
doi={10.1007/s11467-023-1309-z}
}

@article{osterloh2005cold,
  title = {Cold Atoms in Non-Abelian Gauge Potentials: From the Hofstadter "Moth" to Lattice Gauge Theory},
  author = {Osterloh, K. and Baig, M. and Santos, L. and Zoller, P. and Lewenstein, M.},
  journal = {Phys. Rev. Lett.},
  volume = {95},
  issue = {1},
  pages = {010403},
  numpages = {4},
  year = {2005},
  month = {Jun},
  publisher = {American Physical Society},
  doi = {10.1103/PhysRevLett.95.010403}
}

@article{greschner2015density,
  title = {Density-dependent synthetic magnetism for ultracold atoms in optical lattices},
  author = {Greschner, Sebastian and Huerga, Daniel and Sun, Gaoyong and Poletti, Dario and Santos, Luis},
  journal = {Phys. Rev. B},
  volume = {92},
  issue = {11},
  pages = {115120},
  numpages = {10},
  year = {2015},
  month = {Sep},
  publisher = {American Physical Society},
  doi = {10.1103/PhysRevB.92.115120}
}

@article{gorg2019realization,
author={G{\"o}rg, Frederik
and Sandholzer, Kilian
and Minguzzi, Joaqu{\'i}n
and Desbuquois, R{\'e}mi
and Messer, Michael
and Esslinger, Tilman},
title={Realization of density-dependent Peierls phases to engineer quantized gauge fields coupled to ultracold matter},
journal={Nature Physics},
year={2019},
month={Nov},
day={01},
volume={15},
number={11},
pages={1161-1167},
issn={1745-2481},
doi={10.1038/s41567-019-0615-4}
}

\clearpage

\begin{widetext}
\begin{center}
\begin{large}
{\bf Supplemental Material for: Non-Abelian Gauge Field Mechanics}
\end{large}
\end{center}
\end{widetext}

\section*{Overview}

In Section~\ref{sec:theory}, we review the theoretical mapping from our classical experimental systems to the desired Hamiltonian physics. In Section~\ref{sec:expt}, we present details of the experimental implementation, beginning with an overview of the set-up before discussing signal processing, oscillator calibration and finally our measurement protocol for extracting the magnitude of the Wilson loop. In Section~\ref{sec:2D}, we provide additional information to support the analysis of the 2D non-Abelian Hermitian lattice model [Eq.~1 in the main text], beginning with supplemental experimental data for the spectrum of this model, before presenting a more detailed theoretical discussion of non-Abelian gauge fields and the minimal set of loop operators, as used in the main part of the paper. Finally, in Section~\ref{sec:1d}, we include additional supplemental dynamical data for the 1D non-Abelian Hatano-Nelson chain [Eq.~5 in the main text]. 

\section{Theoretical Mapping} \label{sec:theory}

The theoretical models presented in the main text are  realized experimentally in our classical mechanical system by exploiting a mapping from
Newton’s equations of motion onto the Heisenberg equations for a desired target Hamiltonian. A full derivation of this mapping is given for the spinless case in Ref.~\cite{Anandwade-synthetic}. Extending this to realize SU(2) gauge fields is straightforward; we represent each lattice site with two mechanical oscillators, whose relative amplitude and phase encode a local pseudo-spin-$1/2$ degree of freedom. The behavior of each of these oscillators is described by its position and momentum variables $(x_{i,\sigma},p_{i,\sigma})$, with the labels $(i, \sigma)$, where $i$ is the lattice site index and $\sigma=\uparrow,\downarrow$ distinguishes between the two oscillators at a given site. 

For such a set of identical classical harmonic oscillators, Newton's equations of motion can be written as
\begin{equation}
m\dot{x}_{i,\sigma}(t) = p_{i,\sigma}(t), \qquad 
\dot{p}_{i,\sigma}(t) = -m\omega^2 x_{i,\sigma}(t),
\tag{S1}
\end{equation}
where $\omega$ is the natural frequency and $m$ is the mass. For notational simplicity, we introduce the variables
\begin{equation}
\tilde{x}_{i,\sigma} \equiv -\omega^2 x_{i,\sigma}(t),
\qquad
\tilde{p}_{i,\sigma} \equiv -\frac{\omega^2}{m}p_{i,\sigma}(t), \nonumber
\end{equation}
to rewrite the equations of motion as
\begin{equation}
\dot{\tilde{x}}_{i,\sigma} = \tilde{p}_{i,\sigma}, \qquad 
\dot{\tilde{p}}_{i,\sigma} = -\omega^2 \tilde{x}_{i,\sigma}.
\tag{S2}
\end{equation}
In order to simulate the desired spinful model Hamiltonians, we add real-time feedback forces such that these equations become
\begin{equation}
\dot{\tilde{x}}_{i,\sigma} = \tilde{p}_{i,\sigma}, \qquad 
\dot{\tilde{p}}_{i,\sigma} = -\omega^2 \tilde{x}_{i,\sigma} + f_{i,\sigma},
\tag{S3}
\label{S3}
\end{equation}
where $f_{i,\sigma}$ can, in general, depend on all oscillator coordinates and momenta,
\begin{equation}
f_{i,\sigma}
=
f_{i,\sigma}
\left(
\{\tilde{x}_{j,\sigma'},\tilde{p}_{j,\sigma'}\}
\right), \nonumber
\end{equation}
allowing feedback to generate both inter-site hopping and coupling between the two pseudo-spin components.

To map this system onto the Heisenberg equations of motion, we introduce the classical complex amplitudes
\begin{equation}
\alpha_{i,\sigma}
\equiv
\sqrt{\frac{\omega}{2}}\tilde{x}_{i,\sigma}
+
i\sqrt{\frac{1}{2\omega}}\tilde{p}_{i,\sigma},
\tag{S4}
\end{equation}
which are analogous to the annihilation operators associated with quantum harmonic oscillators. Hence, it follows that
\begin{equation}
\tilde{x}_{i,\sigma}
=
\sqrt{\frac{1}{2\omega}}
\left(
\alpha_{i,\sigma}+\alpha_{i,\sigma}^*
\right),
\qquad
\tilde{p}_{i,\sigma}
=
-i\sqrt{\frac{\omega}{2}}
\left(
\alpha_{i,\sigma}-\alpha_{i,\sigma}^*
\right),
\tag{S5}
\end{equation}
such that the equations of motion can be written as
\begin{equation}
\dot{\alpha}_{i,\sigma}
=
-i\omega\alpha_{i,\sigma}
+
\frac{i}{\sqrt{2\omega}}f_{i,\sigma},
\label{S6}
\tag{S6}
\end{equation}
with $f_{i,\sigma}$ now being a function in general of the set of classical amplitudes $\{\alpha_{j,\sigma'},\alpha_{j,\sigma'}^*\}$. (Note that this approach can also be used to engineer other effects, such as on-site gain and losses or classical nonlinearities~\cite{Anandwade-synthetic}.)

Without feedback, the complex amplitudes of each oscillator would evolve independently as $\alpha_{i,\sigma}(t)\propto e^{-i\omega t}$, provided we ignore natural dissipation, which can be assumed to be weak. Including feedback forces modifies the dynamics; however, as long as these forces are sufficiently weak and $\omega$ remains the largest frequency scale in the system, any changes will be small compared to the natural oscillator frequency. In this limit, the Rotating Wave Approximation (RWA) can be applied to re-write Eq.~\ref{S6} as
\begin{equation}
\dot{\alpha}_{i,\sigma}
=
-i\omega\alpha_{i,\sigma}
+
\sum_{j,\sigma'}
\frac{i}{\sqrt{2\omega}}
F^{\mathrm{RWA}}_{i\sigma,j\sigma'}
\alpha_{j,\sigma'},
\label{S7}
\tag{S7}
\end{equation}
where $\sum_{j,\sigma'}F^{\mathrm{RWA}}_{i\sigma,j\sigma'}\alpha_{j,\sigma'}$ corresponds to the applied feedback force obtained under the RWA, i.e. keeping only co-rotating terms. The coefficient $F^{\mathrm{RWA}}_{i\sigma,j\sigma'}$ describes the feedback-induced coupling from oscillator $(j,\sigma')$ to oscillator $(i,\sigma)$. This is then of the same form as the Heisenberg equation of motion for a spinful bosonic tight-binding Hamiltonian (with $\hbar=1$),
\begin{equation}
H
=
\sum_{i,\sigma}
\omega
\hat{c}_{i,\sigma}^{\dagger}
\hat{c}_{i,\sigma}
-
\frac{1}{\sqrt{2\omega}}
\sum_{i,j}
\sum_{\sigma,\sigma'}
\hat{c}_{i,\sigma}^{\dagger}
F^{\mathrm{RWA}}_{i\sigma,j\sigma'}
\hat{c}_{j,\sigma'} .
\tag{S8}
\label{S8}
\end{equation}
Here $\hat{c}_{i,\sigma}^{\dagger},\hat{c}_{i,\sigma}$ are, respectively, bosonic creation and annihilation operators for the pseudospin state $\sigma$ on the $i$-th lattice site. When considering non-Abelian gauge fields, it is helpful to group the two operators at each lattice site into a spinor $\hat{c}_i=(\hat{c}_{i,\uparrow},\hat{c}_{i,\downarrow})^T$ as we did in the main text. Then the feedback-induced couplings between sites $j$ and $i$ can be viewed as $2\times2$ matrices acting on the pseudospin degree of freedom,
\begin{equation}
{F}^{\mathrm{RWA}}_{ij}
=
\begin{pmatrix}
{F}^{\mathrm{RWA}}_{i\uparrow,j\uparrow} &
{F}^{\mathrm{RWA}}_{i\uparrow,j\downarrow} \\
{F}^{\mathrm{RWA}}_{i\downarrow,j\uparrow} &
{F}^{\mathrm{RWA}}_{i\downarrow,j\downarrow}
\end{pmatrix}. \nonumber
\end{equation}
Experimentally, all of our desired spin-dependent coupling terms can be realized by applying suitable position- and momentum-dependent feedback. To see this explicitly, we write a general matrix element of the RWA feedback force as
\begin{equation}
F^{\mathrm{RWA}}_{i\sigma,j\sigma'}
=
F^R_{i\sigma,j\sigma'}
+
iF^I_{i\sigma,j\sigma'} , \nonumber
\end{equation}
where $F^R_{i\sigma,j\sigma'}$ and $F^I_{i\sigma,j\sigma'}$ are real. The desired co-rotating term $F^{\mathrm{RWA}}_{i\sigma,j\sigma'}\alpha_{j,\sigma'}$ in Eq.~\ref{S7} is therefore generated via Eq.~\ref{S3} by applying to oscillator $(i,\sigma)$ a real feedback force of the form
\begin{equation}
f_{i,\sigma}
=
\sum_{j,\sigma'}
\left[
A_{i\sigma,j\sigma'} \tilde{x}_{j,\sigma'}
+
B_{i\sigma,j\sigma'} \tilde{p}_{j,\sigma'}
\right], \nonumber 
\end{equation}
where $A_{i\sigma,j\sigma'}= \sqrt{2 \omega} F^R_{i\sigma,j\sigma'}$ and $B_{i\sigma,j\sigma'} = - \sqrt{2/\omega} F^I_{i\sigma,j\sigma'}$. 
Thus, the real part of the desired RWA coupling is implemented using position-dependent feedback, while the imaginary part is implemented using momentum-dependent feedback.

As an example, consider the 1D non-Abelian Hatano-Nelson model studied in the main text,
\begin{eqnarray} H = \sum_m \left[J_L \hat{c}^\dagger_m e^{i \theta_L \sigma_y} \hat{c}_{m+1} + J_R \hat{c}^\dagger_{m+1} e^{i \theta_R \sigma_x} \hat{c}_{m} \right] 
\nonumber
\end{eqnarray} 
where $m$ is the site index along the $x$ direction. The corresponding feedback-force matrix for the coupling to site $m$ from neighboring site $m+1$ is then 
\begin{equation} 
{F}^{\mathrm{RWA}}_{m,m+1} = -\sqrt{2\omega}\, J_L \begin{pmatrix} \cos\theta_L & \sin\theta_L \\ -\sin\theta_L & \cos\theta_L \end{pmatrix}, \nonumber
\end{equation} 
while the coupling to site $m+1$ from site $m$ is 
\begin{equation} 
{F}^{\mathrm{RWA}}_{m+1,m} = -\sqrt{2\omega}\, J_R \begin{pmatrix} \cos\theta_R & i\sin\theta_R \\ i\sin\theta_R & \cos\theta_R \end{pmatrix}. \nonumber
\end{equation} 
Note that, unlike in a Hermitian system, these two feedback matrices are not constrained to be Hermitian conjugates, reflecting that the hopping is non-reciprocal. However, this is not an issue for our experimental set-up, and we are able to easily realize such non-Hermitian effects by applying different feedback forces to oscillators on different sites. More specifically, for the left-moving hopping matrix $F^{\mathrm{RWA}}_{m,m+1}$, all entries are real; hence, this link requires only position-dependent feedback
\begin{align}
f_{m,\uparrow}^{(m+1)}
&\propto
-\sqrt{2\omega}J_L
\left[
\cos\theta_L\, \tilde{x}_{m+1,\uparrow}
+
\sin\theta_L\, \tilde{x}_{m+1,\downarrow}
\right], \nonumber
\\
f_{m,\downarrow}^{(m+1)}
&\propto
-\sqrt{2\omega}J_L
\left[
-\sin\theta_L\, \tilde{x}_{m+1,\uparrow}
+
\cos\theta_L\, \tilde{x}_{m+1,\downarrow}
\right],\nonumber
\end{align}
where $f_{m,\sigma}^{(m+1)}$ denotes the part of the force applied to the oscillator indexed by $(m, \sigma)$, which is due to the two oscillators at site $(m+1)$. 
For the right-moving hopping matrix $F^{\mathrm{RWA}}_{m+1,m}$, the diagonal entries are real while the off-diagonal entries are imaginary. The corresponding feedback forces are therefore instead
\begin{align}
f_{m+1,\uparrow}^{(m)}
&\propto
-\sqrt{2\omega}J_R
\left[
\cos\theta_R\, \tilde{x}_{m,\uparrow}
+
\sin\theta_R\, \tilde{p}_{m,\downarrow}
\right], \nonumber
\\
f_{m+1,\downarrow}^{(m)}
&\propto
-\sqrt{2\omega}J_R
\left[
\sin\theta_R\, \tilde{p}_{m,\uparrow}
+
\cos\theta_R\, \tilde{x}_{m,\downarrow}
\right].\nonumber
\end{align}
The same procedure can straightforwardly be carried out for the other two models studied in the main text to determine the corresponding feedback forces in our experiment. 

\section{Experimental Implementation} \label{sec:expt}

\subsection{Set-up Overview}

In our experiment, each mechanical oscillator consists of a 3D printed rounded rectangular prism, suspended from a 3D printed support structure by two steel springs. The pair of springs work in tandem to give a uniform restoring force for both positive and negative displacements along the oscillator axis. The springs are attached to the main mass by alternately nylon or metal screws, allowing for gross adjustment of the oscillator frequency by replacing or removing a screw as this will alter the oscillator's mass. We take the first longitudinal mode (along the vertical axis) as our oscillator’s primary mode and neglect any transverse or torsional modes as they are well-separated in frequency from the fundamental mode.

In order to measure the state of the oscillator, we embed a solid-state analog two-axis accelerometer evaluation board (EVAL-ADXL203) in the main oscillator body. In our
experiment, we use only one axis of the accelerometer. The evaluation board comes with capacitors pre-installed that limit the bandwidth to 50 Hz for low-noise operation; this is sufficient for our needs as our oscillators’ natural frequency is $\omega / 2\pi \approx 13.06$~Hz. This signal is then sampled and digitized at 1 kHz by a National Instruments PXIe system analog-to-digital conversion card via high-gauge wire connection.

We apply the feedback force to each oscillator using a combination of a small dipole magnet, embedded in the main mass, and a pair of anti-Helmholtz coils attached to the support structure. We control the current in these coils with a laser current driver, with only one direction of current flow allowed~\cite{velkovsky2024observation}. This creates an axial magnetic field gradient, which for small displacements gives a linear force on the oscillator. We control the current driver output using the same PXIe interface as our data readout, providing a single input-output interface for the experiment. 

Our current driver is a Wavelength Electronics FL591FL evaluation board, designed as a laser diode current driver. It has two independent outputs that can each drive 250 mA of current, which can be combined into a single 500 mA current driver output. The current output is set by a 0-2 V signal, which we drive directly from the computer to set the force. We operate the boards in constant current mode, which directly ties the current output to the voltage signal, and use a single driver per oscillator in single-output mode for higher dynamic range.

\subsection{Signal Processing}

As described in the previous section, our primary signal for the experiment consists of a real-time measurement of each oscillator’s acceleration $a(t)$ as an analog voltage signal. We sample this signal at 1 kHz (i.e. $\Delta t = 1$~ms) and read it into a National Instruments LabVIEW program using the built-in NI interface for our ADC. Given the timescales of our experiment, this gives essentially a continuous measurement of the acceleration. We digitally filter the signal in order to reduce high-frequency noise in the system. We then numerically differentiate the acceleration to obtain the oscillators’ jerk $j(t)\equiv \dot{a}(t)$. Because $a(t) \propto x(t)$ and $j(t) \propto p(t)$ for a harmonic oscillator, we treat the $a$ and $j$ signals as proxies for the position $x$ and momentum $p$. We additionally multiply (i.e., normalize) the signals by an empirically determined constant to put them on a common scale, reflecting the equipartition of energy in harmonic oscillators. Due to the phase lag introduced by our filtering and differentiation, the two signals are rotated to produce pure $x$ and $p$ proxies. This rotation is determined experimentally, based on the oscillators' response to $x$- and $p$-dependent self-feedback.

After the initial signal processing, the current time in experiment, the current iteration (for looped or list run experiments), and the $x$ and $p$ signals for all the oscillators are combined and fed into a LabVIEW MathScript code block. This block contains user-defined code that further calibrates the signals and runs the main portion of the experiment. Further discussion of calibrations are included below. The code computes the necessary feedback forces based on the input signals and then outputs these to the current driver. In parallel, the $x$ and $p$ signals are recorded in binary format for later analysis.

Within the code block, the experimental run consists of three phases. In the first phase, the oscillators are excited to produce an initial state by driving the sites with a high-amplitude sine wave for two seconds. Following that, the main experiment runs, with the force for each oscillator computed from the input signals as detailed above. The final phase of the run damps out the motion of all of the oscillators, either to end the experiment or in preparation for the next run.

This experimental protocol is highly versatile as it can allow us to simulate a wide-range of different Hamiltonians including arbitrary connectivity, non-Hermitian effects, time-dependent Hamiltonians and mean-field nonlinearities~\cite{Anandwade-synthetic}. However, there are a few limitations on the experimental system, namely in keeping the size of the excitations small (to preserve linearity of the natural spring forces) and computational complexity of the given Hamiltonian. In the current setup, the full code block needs to execute before the next sample of the input signals is taken. Therefore, if the output forces take too long to compute, it can cause glitches and unwanted behavior in the experiment. Also, if any of the oscillators’ amplitudes exceeds a preset value, the iteration ends and all oscillators are damped out to prevent damage to the system. 

\begin{figure*}[t!]
	\includegraphics[width=1.85\columnwidth]{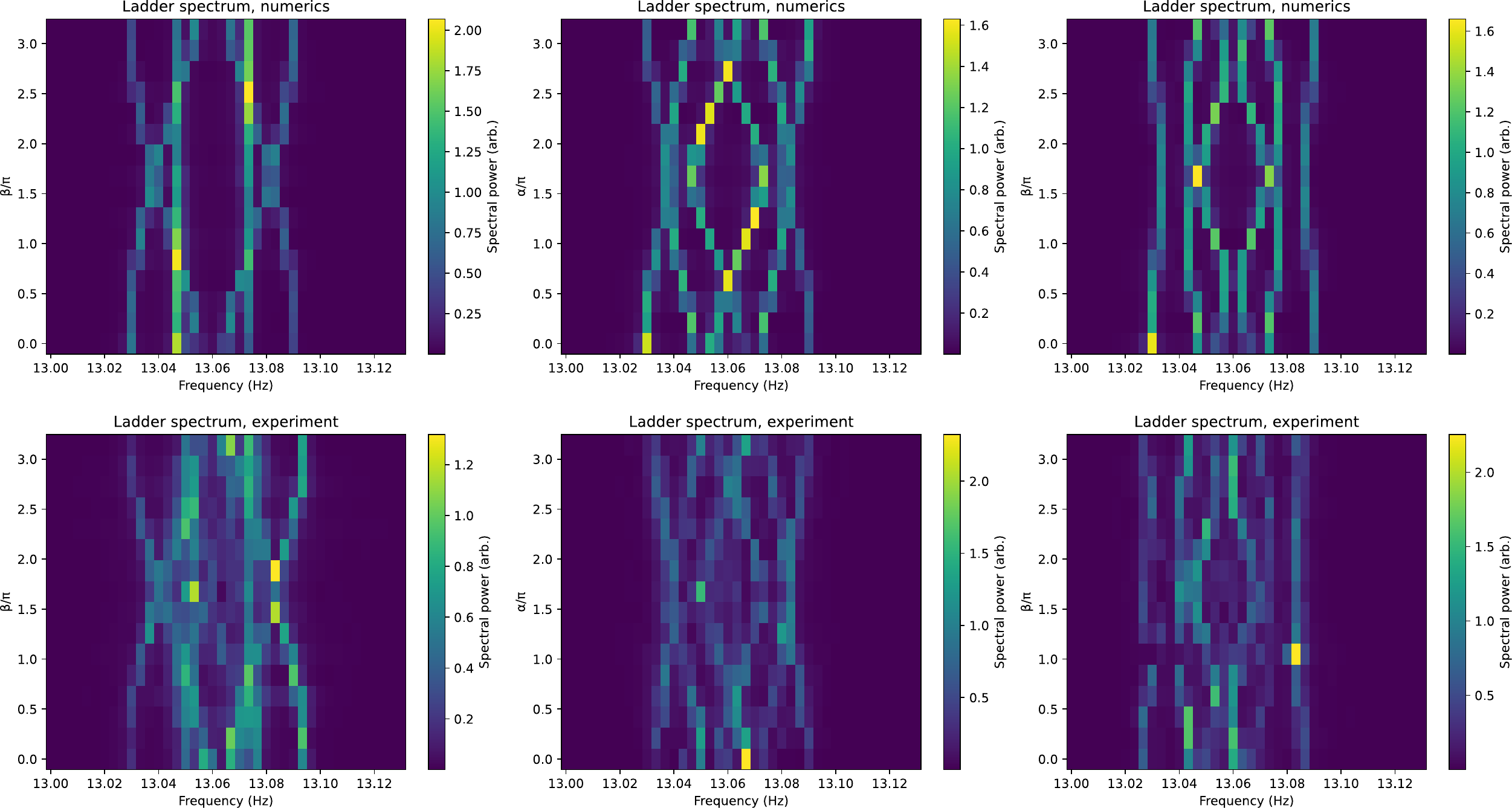}
	\centering
	\caption{\label{FIG:NA Spect}
		The energy spectrum of our 6-site non-Abelian Hermitian ladder [c.f. Eq.~1 in the main text], as extracted using a ringdown measurement observing 300~s of dynamics. In each case, we excite the spin-up bottom-left site in the lattice, before allowing the system to dynamically evolve. We then take the Fourier power spectrum of each oscillator’s position $x$ versus experimental time and sum to obtain the full spectrum of the system.  Numerically simulated and experimentally measured spectrum, respectively, for (a,b) varying $\beta$ with $\alpha= \pi/2$, (c,d) varying $\alpha$ with $\beta= \pi/2$, (e,f) varying $\alpha$ with $\beta = 5\pi/16$.
	}
\end{figure*}

\subsection{Calibration of oscillators}

To control for errors and ensure that our simulations are as accurate as possible, we perform several calibration steps on the oscillators. First, we subtract off any voltage offset from the signal to center it on zero. We determine the scaling factor from an initial excitation of the oscillator, using the average peak-to-peak height. The rotation angle is determined via feedback calibration. We first take a ringdown measurement with no feedback to calculate each oscillator’s natural damping coefficient. Then, we apply a self-feedback force proportional to the position $x$ and observe the ringdown. By our assumptions, feedback on $x$ should only affect the oscillator’s frequency, while feedback on $p$ should only affect the damping. We tune the rotation angle to minimize the effect of $x$ feedback on the damping factor.

Next, we correct for non-ideal behavior on the individual oscillator level. As mentioned previously, self-feedback on the oscillator’s momentum affects the on-site damping (equivalent to an imaginary diagonal matrix element). We take successive ringdown measurements while tuning the momentum self-feedback until the oscillator has little to no loss or gain. Using this method, we can keep the oscillator excited at a constant level on the order of 500~s. We then calibrate the frequency of the oscillators to reduce disorder across the array. We tune the frequency coarsely using mass adjustments, and tune it finely using position-dependent self-feedback. Specifically, we perform a virtual homodyne measurement, comparing each oscillator’s phase to a reference signal in code, and tune a self-feedback coefficient until the oscillator’s phase is nearly constant with respect to the reference.

The experimental hopping rates $J$ are calibrated based on direct measurement of two-site tunneling dynamics, and our experiments all operate with approximately uniform hopping rates of $J/2\pi \approx 32$~mHz.

\subsection{Wilson Loop Measurement Protocol}

In the main text, we demonstrate the genuine non-Abelian nature~\cite{SoljacicNA} of our system through direct measurement of the magnitude of the Wilson loop. We extract this information from four measurements in a procedure inspired by quantum process tomography~\cite{QuantumProcessTomography}. In our procedure, we initialize the system by injecting energy into a single lattice site in one of four defined states
\begin{equation}
 \ket{0} \equiv \ket{\uparrow},
  \ket{1} \equiv \ket{\downarrow},
  \ket{2} \equiv \frac{(\ket{\uparrow} + \ket{\downarrow})}{\sqrt{2}},
  \ket{3} \equiv \frac{(\ket{\uparrow} - i\ket{\downarrow})}{\sqrt{2}}, \tag{S9}
 \label{eq:NAStates}
\end{equation}
which we achieve by exciting the two oscillators representing that site with the appropriate phase relationship. We then turn on a single coupling link for exactly one tunneling time, calibrated against our empirically-measured experimental tunneling rate. This coupling transfers the state to the next site while performing the appropriate state-dependent rotation in the internal subspace [c.f. Eqs.~1 and 5 in the main text]. We repeat this procedure until the excitation returns to the original site, at which point we have performed a single loop. To note, while the excitation is traversing the loop, we intentionally damp out remnant amplitude at the oscillators of the original site to avoid interference effects upon the completion of the full loop.

After the excitation traverses the full loop, we characterize the final state by three real parameters $(a_i, b_i, \varphi_i)$:
\begin{align}
	\ket{f_i} = a_i\ket{\uparrow} + b_ie^{i\varphi_i}\ket{\downarrow} , 
	\label{eq:NAFinalState} \nonumber
\end{align}
where $i=0,1,2,3$ denotes the index of the corresponding starting state from Eq.~\ref{eq:NAStates}. For a given $(\alpha,\beta)$ pair, we can then combine the four sets of final-state parameters to calculate the intermediate values,
\begin{align}
	\begin{split}
	A &= a_0^2 ,  \qquad
	B = b_1^2,
	\end{split} \nonumber \\
	\begin{split}
	C &= a_2b_2e^{i\varphi_2} - i a_3b_3e^{i\varphi_3}-\frac{1-i}{2}(a_0b_0e^{i\varphi_0}+a_1b_1e^{i\varphi_1}),
	\end{split} \nonumber\\
	\begin{split}
	D &= a_2 b_2 e^{-i\varphi_2} + a_3 b_3 e^{-i\varphi_3}-\frac{1+i}{2}(a_0 b_0 e^{-i\varphi_0}+a_1 b_1 e^{-i\varphi_1}),
	\end{split} \nonumber
\end{align}
from which the magnitude of the Wilson loop is given by,
\begin{equation}
	|W_i| = \sqrt{|\Re(A + B + C + D)|} . \tag{S10}
	\label{eq:TraceWL}
\end{equation}
This procedure has been used to obtain the experimental results shown in Figs.~1 and 2 in the main text.

\section{2D non-Abelian Hermitian Lattice Model} \label{sec:2D}

\subsection{Measurement of Spectrum}

To show that we have engineered the desired 2D non-Abelian lattice model, we experimentally measure the spectra of our small lattice as a function of the non-Abelian gauge parameters $\alpha$ and $\beta$ [c.f. Eq.~1 in the main text].  Experimentally, each spectrum is obtained by observing the dynamics over 300~s in a ringdown measurement. To do this, we excite a single oscillator (chosen as the spin-up bottom-left site) and then allow the system to evolve under the 6-site ladder Hamiltonian (i.e. with all coupling links turned on). Numerical simulations suggest that this initial condition is sufficient to project onto all of the eigenstates of the system and thus observe the energy spectrum of the ladder. We extract the spectral peaks in frequency by taking the summed Fourier power spectrum of the $x$ dynamics at each lattice site. We then restrict our attention to a small region around the oscillators’ natural frequency, which in our experiment is $\omega / 2\pi \approx 13.06$~Hz. The resulting spectra are shown for three different parameter cuts in Fig.~\ref{FIG:NA Spect}, where in each case, we also compare our results to a numerically simulated version of the experiment using the experimentally-determined tunneling rate and natural oscillator frequency. In all three spectral sweeps, we see good agreement between the numerics and the measured spectra in the number, spacing, and ($\alpha$, $\beta$) dependence of the modes.

\begin{figure}[t!]
	\includegraphics[width=0.85\columnwidth]{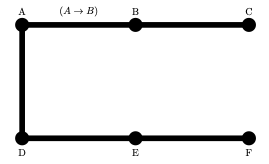}
	\centering
	\caption{\label{FIG:spanning}
		A choice of spanning tree for our 6-site mechanical lattice. We have labelled the vertices/lattice sites as A, B, C, D, E and F. All vertices are connected through nearest-neighbour edges, such that there is a path joining every vertex to every other vertex but without loops in the system. As an example, we have labelled one edge as $(A \rightarrow B)$ indicating the (directional) edge between vertices $A$ and $B$. 
        }
\end{figure}

\subsection{Genuine Non-Abelian Gauge Fields}

In the main text, we use the magnitude of different Wilson loops in Fig.~1 to draw conclusions about when the gauge field is Abelian or non-Abelian for different parameter values. In this section, we shall explain in more detail the reasoning behind these inferences. 

As reviewed in the main text, a gauge field is said to be geninuely non-Abelian if and only there is at least one set of loop operators which do not commute, i.e. $[U_i,U_j] \neq 0$ for some $i,j$ with $U_i$ being a loop operator indexed by $i$. Note that we assume all $U_i$ have the same base point (start and end point). In general, this implies that proving a gauge field is Abelian requires us to check the commutation of every possible loop operator in the system. However, for a given system size, it can be possible to identify a basis set of loop operators (for each given base point), such that every other loop can always be expressed as a product of these operators. For our system, if we choose the base point as the bottom left lattice site in Fig.~1 in the main text, then a suitable basis corresponds to the four loop operators, $U_1$, $U_2$ and their Hermitian conjugates; as the justification of this point is technical, we leave it to the following section. 

Another useful result is that, in our system, all loop operators, $U_i$, are in the group SU(2); this means that each loop operator is unitary with $\text{det}(U_i)= 1$. Consequently, all the loop operators have two eigenvalues, $\lambda_1$ and $\lambda_2$, which are constrained to the unit circle and are complex conjugates of each other, i.e. $\lambda_1 = e^{i\theta_i}$ and $\lambda_2 = e^{-i\theta_i}$ for the loop operator $U_i$. This allows us to always write 
\begin{equation*}
    W_i \equiv  tr(U_i) = e^{i\theta_i} + e^{-i\theta_i} = 2\cos{(\theta_i)}.
\end{equation*}
For the special values of $|W_i| = 2$, we can see that $\cos{(\theta_i)} = \pm 1$ which coincides with our eigenvalues being both either $1$ or both $-1$. There are only two choices of matrices in SU(2) satisfying this, corresponding to $\mathbf{1}$ or $-\mathbf{1}$ respectively; hence we can conclude that the loop operator $U_i \propto \mathbf{1}$ if we measure $|W_i| = 2$.

Putting these facts together, let us now consider the four different Wilson loops discussed in the main text for the 2D non-Abelian Hermitian model [c.f. Fig.~1]. Firstly, if $|W_1|\equiv |tr(U_1)| = 2$, then it follows $U_1 \propto \mathbf{1}$. This means that in our basis the only non-trivial remaining loop operators are $U_2$ and $U_2^\dag$, and as all possible loop operators can be expressed as products of these two matrices, they all commute with each other. Hence, we can conclude that the gauge field is Abelian wherever we measure $|W_1|=2$. Secondly, identical reasoning applies to the case when $|W_2|\equiv |tr(U_2)| = 2$, except that now the only non-trivial loop operators in our basis are $U_1$ and $U_1^\dag$. Thirdly, if $|W_3|\equiv |tr(U_2U_1)| = 2$, then $U_2U_1 \propto \mathbf{1}$. This in turn means that $[U_2,U_1] = 0$ and thus all the basis loop operators $U_1$, $U_2$, $U_1^\dag$ and $U_2^\dag$ commute, again proving that the gauge field must be Abelian. Finally, if $|W_4|\equiv |tr(U_2^\dag U_1^\dag U_2 U_1)| = 2$, then $U_2^\dag U_1^\dag U_2U_1 = \pm \mathbf{1}$. This implies that $U_2U_1 = \pm U_1U_2$, but this is not strong enough to distinguish between Abelian and non-Abelian gauge fields. However, if we instead consider $|W_4|\equiv|tr(U_2^\dag U_1^\dag U_2U_1)| \neq 2$, then we can conclude that $U_2U_1 \neq  U_1U_2$ must be true, i.e. the loop operators do not commute and our gauge field must be genuinely non-Abelian.

\subsection{Derivation of basis loops}\label{subsec:basisloops}

We wish to establish the existence of a minimal set of loop operators in our finite 2D non-Abelian lattice model [c.f. Eq.~1 in the main text], such that all loops can be expressed as products of this basis set. To do so, we begin by employing a concept from graph theory known as a spanning tree. This is constructed starting from the vertices or lattice sites of the graph; for our model, there are six sites, which we label as A, B, C, D, E, and F respectively. We then add edges to the graph such that all vertices are connected but without any loops, as shown in Fig.~\ref{FIG:spanning} for one choice of the spanning tree where we only include nearest-neighbor edges.
Each edge can be traversed in two opposite directions, 
$(V_i \rightarrow V_j$) and $(V_j \rightarrow V_i$), where $V_{i}$ and $V_{j}$ correspond to the corresponding pair of nearest-neighbor vertices. As our model is Hermitian [Eq.~1 in the main text], these have the property that $(V_i \rightarrow V_j) (V_j \rightarrow V_i) = \mathbf{1}$. 

From this construction, we can define a unique path from any vertex to any other vertex in the spanning tree. Hence, we introduce a function,
\begin{equation}
F_\alpha(V_i\rightarrow V_j) = P^{-1}_\alpha(V_j) (V_i\rightarrow V_j) P_\alpha(V_i)
\tag{S11}
\label{eq:s11}
\end{equation}
where $\alpha$ is some reference vertex, and $P_{\alpha}(V_i)$ is the unique path from $\alpha$ to vertex $V_i$ as defined by our choice of spanning tree. For cases like $P_{\text{A}}(B)$, where $A$ and $B$ are nearest-neighbor vertices, the unique path is simply      
\begin{equation}
    P_{\text{A}}(\text{B}) = (\text{A}\rightarrow \text{B}). \nonumber
\end{equation}
The $F_\alpha (V_i\rightarrow V_j)$ function introduced above has the interesting property that if the direct connection,  $(V_i\rightarrow V_j)$, is in the spanning tree then the function will simplify to the identity, whereas if it is not, then the function will output a generic operator. For example, consider a case where the direct connection is present as
\begin{align*}
    F_{\text{A}}(\text{B}\rightarrow\text{C}) =& P^{-1}_{\text{A}}(\text{C})(\text{B}\rightarrow \text{C}) P_{\text{A}}(\text{B}) \\
    =&(\text{B}\rightarrow \text{A}) (\text{C}\rightarrow \text{B}) (\text{B} \rightarrow \text{C}) (\text{A}\rightarrow \text{B}) \\
    =& \mathbf{1}.
\end{align*}
It can be checked that this holds similarly for all other edges present in the spanning tree. 
However, if we instead consider a direct connection that is not present, e.g.
\begin{align*}
    F_{\text{A}}(\text{B}\rightarrow\text{E}) =& P^{-1}_{\text{A}}(\text{E})(\text{B}\rightarrow \text{E}) P_{\text{A}}(\text{B}) \\
    =&(\text{D}\rightarrow \text{A}) (\text{E}\rightarrow \text{D}) (\text{B} \rightarrow \text{E}) (\text{A}\rightarrow \text{B}) ,
\end{align*}
then this expression does not reduce to the identity for every choice of matrices representing the edges. In our small lattice, the only other nearest-neighbor connection that is not present in the spanning tree is $(C \rightarrow F)$, which generates an analogous but generically different function, $F_A (C \rightarrow F)$. Note that $F_A (E \rightarrow B)$ and $F_A (F \rightarrow C)$ will be related to the corresponding functions above by Hermitian conjugation as our model is Hermitian. 

Now, let us consider all loops starting and ending at the vertex $A$ (where the choice of $A$ is arbitrary). We can define each loop as
\begin{align}
    \gamma_A &= \prod_{i=0}^{n-1} (V_{i}\rightarrow V_{i+1}) \nonumber \\
    &= (V_{n-1} \rightarrow V_{n}) ... (V_{1} \rightarrow V_{2})(V_{0} \rightarrow V_{1}), \nonumber
\end{align}
where $n\!-\!1$ is the total number of edges in the loop, $V_{j}$ represents the $j^{\text{th}}$ vertex visited, and $V_{0} = V_{n} = \text{A}$. Hence, using $P_{\text{A}}(V_{n}) = P_{\text{A}}(V_{0}) = P_{\text{A}}(A)= \mathbf{1}$ and by inserting $P^{-1}_{\text{A}}(V_j)P_{\text{A}}(V_j) = \mathbf{1}$, we can turn the previous expression into the same form as Eq.~\ref{eq:s11} as shown here 
\begin{align}
    \gamma_A =& 
     \prod_{i=0}^{n-1} P^{-1}_{\text{A}}(V_{i+1})(V_{i}\rightarrow V_{i+1})P_{\text{A}}(V_{i}) \nonumber \\ 
    =& \prod_{i=0}^{n-1} F_{\text{A}}(V_{i}\rightarrow V_{i+1}). \nonumber
\end{align}
This allows us to rewrite any loop with a base point at vertex A as a product of functions, which are only non-trivial if the direct connection in the function's argument is not present in the spanning tree, as discussed above. For our spanning tree, this allows us to write all loops as products of the only four non-trivial functions: $F_{\text{A}}(\text{B}\rightarrow \text{E})$, $F_{\text{A}}(\text{C}\rightarrow \text{F})$, $F_{\text{A}}(\text{E}\rightarrow \text{B})$ and $F_{\text{A}}(\text{F}\rightarrow \text{C})$. As our  Hamiltonian is Hermitian, this is equivalent to only considering $F_{\text{A}}(\text{B}\rightarrow \text{E})$, $F_{\text{A}}(\text{C}\rightarrow \text{F})$ and their Hermitian conjugates.

\begin{figure*}[t!]
	\includegraphics[width=1.85\columnwidth]{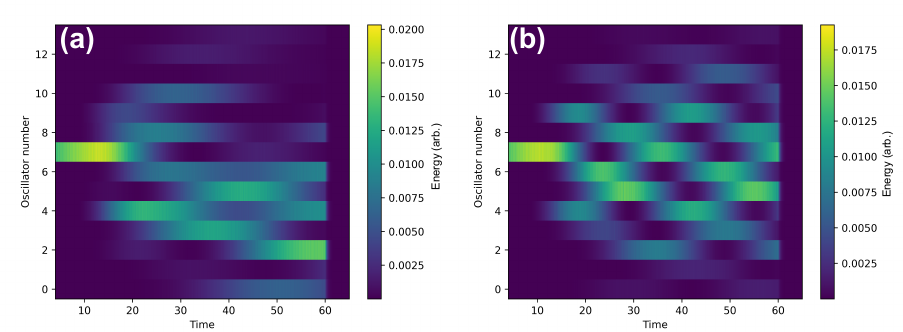}
	\centering
	\caption{\label{FIG:1ddynamics}
		Experimentally-measured dynamics in the 1D non-Abelian Hatano-Nelson chain, starting from an initial excitation of the middle lattice site ($m=4$) in the ``spin down" configuration (oscillator number 8), for (a) $(\theta_R, \theta_L) = (\frac{10 \pi}{16},\frac{4 \pi}{16})$ and (b) $(\theta_R, \theta_L) = (\frac{8 \pi}{16},0)$. The latter corresponds to parameter values for which we expect a bias [see Fig. 3 in the main text], reflected here by a shift dynamically towards the left-hand side of the chain. The former corresponds to values for which we do not expect a bias, reflected here by the excitation weight remaining near the middle of the chain at long times. 
        }
\end{figure*}

Finally, we can connect this construction to the loop operators used in the
main text and in the preceding section. For the spanning tree shown in
Fig.~\ref{FIG:spanning}, the two edges not contained in the
tree are precisely those which close the elementary loops denoted
by $U_1$ and $U_2$ above and in Fig.~1 of the main text. Explicitly, we can then write
\begin{equation}
    U_1 \equiv F_{\rm A}({\rm E}\rightarrow{\rm B}), 
    \qquad
    U_2 \equiv F_{\rm A}({\rm F}\rightarrow{\rm C}), \nonumber
\end{equation}
with the oppositely oriented loops corresponding to
$U_1^\dagger$ and $U_2^\dagger$, respectively. Thus the four non-trivial
functions of the spanning tree are the
four loop operators $U_1$, $U_2$, $U_1^\dagger$ and $U_2^\dagger$ used previously. In other words, any loop operator based at A can be written as a product,
of $U_1,U_2,U_1^\dagger,U_2^\dagger$. Since
$U_i^\dagger=U_i^{-1}$ for our Hermitian model, commutation of $U_1$ and
$U_2$ immediately implies commutation of their inverses and of any product
formed from them. Conversely, if $[U_1,U_2]\neq 0$, then this is a non-Abelian gauge field as two loop operators do not commute. Hence analyzing this basis is sufficient to allow us to diagnose whether our finite system is Abelian or genuinely non-Abelian. 

\section{1D Non-Abelian Hatano-Nelson Dynamics} \label{sec:1d}

As described in the main text, we extract an experimental measure of the left-right eigenstate localization bias for the 1D Non-Abelian Hatano-Nelson [c.f. Fig. 3 in the main text] from observing the dynamics after an initial excitation of the system. Example experimental measurements of the full dynamics are shown in Fig.~\ref{FIG:1ddynamics} for (a) $(\theta_R, \theta_L) = (\frac{10 \pi}{16},\frac{4 \pi}{16})$ and (b) $(\theta_R, \theta_L) = (\frac{8 \pi}{16},0)$. In each case, we excite the system at $t=0$ by injecting energy into oscillator number 8, corresponding to the spin-down oscillator at lattice site $m=4$ by our labeling convention. To note, the oscillator indices as shown in Fig.~\ref{FIG:1ddynamics} are such that lower oscillator indices relate to higher site indices, with oscillators 0 and 1 relating to site 7 (i.e., with panel (a) showing rightward energy flow).

After initialization at the central site, we let the system evolve for 60~s under the Hamiltonian in Eq.~5 in the main text. As described in the main text, we also stabilize the total energy of the system during the evolution to prevent unbounded growth of the oscillators’ amplitudes. Specifically, we apply a uniform (equal strength pre-factor for all of the oscillators) momentum-dependent feedback force that is scaled by the deviation of the total energy from a set target value. At the effective Hamiltonian level, the applied feedback has the form
\begin{eqnarray} H_\textrm{stab} = -i g D \sum_{j,\sigma} \hat{c}^\dagger_{j,\sigma} \hat{c}_{j,\sigma} \ 
\end{eqnarray} 
where $D = (\sum_{m,\sigma} |\psi_{m,\sigma}|^2 - 1)$ is the deviation of the summed squared amplitudes from the target value of 1 and $g$ is a tuning factor. This nonlinear feedback term stabilizes the total energy but does not affect the normalized energy distribution in space. In the experiment, it is turned on after the initial excitation period and before the forces relating to the effective model under study are quenched on, it is kept on throughout the dynamics as shown in Fig.~\ref{FIG:1ddynamics}, and it is turned off at $T = 60$~s at the same time that we apply a large damping force to all of the oscillators, so as to reset their energies to zero in preparation for the subsequent experimental run. We have confirmed in numerical simulations that the applied weak energy-stabilizing feedback does not influence the dynamics of the normalized energy, while it allows us to stay within the linear, small-amplitude regime of our oscillators for practical considerations (to avoid damage and large physical nonlinearities).

As can be seen directly from the real-space dynamics in Fig.~\ref{FIG:1ddynamics}, certain combinations of $\theta_{L/R}$ can induce directional flow of the normalized energy at long times. On the one hand, the system exhibits a clear directional flow towards the left-hand side of the chain (low oscillator number) for panel (a), where we would expect a significant left-right bias (see Fig.~3 in the main text). On the other hand, the excitation weight remains close to the middle of the chain in panel (b), where no significant left-right bias is expected.

Along with the different behavior in panels (a) and (b) of Fig.~\ref{FIG:1ddynamics} with respect to the directional flow of energy, there is also a notable difference in the frequency of oscillatory energy dynamics stemming from beating between system eigenstates. The contrast between slow and fast dynamics in panels (a) and (b) of Fig.~\ref{FIG:1ddynamics}, respectively, reflects the dependence of the real part of the system's energy spectra on the tuning angles $\theta_R$ and $\theta_L$. As shown in Fig.~3 of the main text, the system's energy spectrum is modified as the point gap topology changes with the variation of the non-Abelian tuning angles $\theta_R$ and $\theta_L$, and a pronounced slow-down of the oscillatory energy dynamics is observed when $\theta_R + \theta_L \approx \pi$.
\end{document}